\documentclass[aps,prd,twocolumn,floatfix,nofootinbib,showpacs,superscriptaddress]{revtex4-1}

\usepackage[mathscr,scaled=1.15]{urwchancal}
\DeclareFontFamily{OT1}{pzc}{}
\DeclareFontShape{OT1}{pzc}{m}{it}%
{<-> s * [1.15] pzcmi7t}{}
\DeclareMathAlphabet{\mathpzc}{OT1}{pzc}{m}{it}

\usepackage{amsmath,amsfonts,amssymb,bm}

\usepackage{graphicx}

\usepackage{color}

\usepackage{slashed}

\definecolor{purple}{rgb}{0.5,0,0.5}
\definecolor{blue}{rgb}{0.0,0,0.9}

\begin{document}

\title{Practical scheme from QCD to phenomena via Dyson-Schwinger equations }

\author{Can Tang}
\affiliation{Department of Physics and State Key Laboratory of Nuclear Physics and Technology,
Peking University, Beijing 100871, China}
\affiliation{Collaborative Innovation Center of Quantum Matter, Beijing 100871, China}

\author{Fei Gao}
\affiliation{Department of Theoretical Physics and IFIC, University of Valencia and CSIC, E-46100, Valencia, Spain}

\author{Yu-xin Liu }
\email[Corresponding author: ]{yxliu@pku.edu.cn}
\affiliation{Department of Physics and State Key Laboratory of Nuclear Physics and Technology,
Peking University, Beijing 100871, China}
\affiliation{Collaborative Innovation Center of Quantum Matter, Beijing 100871, China}
\affiliation{Center for High Energy Physics, Peking University, Beijing 100871, China}

\date{\today}

\begin{abstract}
We deliver a scheme to compute the quark propagator and the quark-gluon interaction vertex through the coupled Dyson-Schwinger equations (DSEs) of QCD.
We take the three-gluon vertex into account in our calculations,
and implement the gluon propagator and the running coupling function fitted by the solutions of their respective DSEs.
We obtain the momentum and current mass dependence of the quark propagator and the quark-gluon vertex,
and the chiral quark condensate
%
%
which agrees with previous results excellently.
We also compute the quark-photon vertex within this scheme and give the anomalous chromo- and electro-magnetic moment of quark. The obtained results are consistent with previous ones very well.
These applications manifest that the scheme is realistic
and then practical for explaining the QCD-related phenomena.
\end{abstract}

\maketitle

\section{Introduction}

People have made a lot of efforts on studying the non-perturbative phenomena of QCD,
such as the dynamical chiral symmetry breaking (DCSB), confinement and the low-lying hadron spectra.
As a  basic and important ingredient of QCD, the quark propagator is tightly connected with these phenomena. Among all the studies, the non-perturbative continuum QCD approach  based on the infinite tower of Dyson-Schwinger equations (DSEs) has made fruitful achievements (see, e.g., Refs.~\cite{RW,MR1997,MTR199903,Ptech,Alkofer,BP1,PW1,Munczek,BCPR,BCPQR,ABP1,AP,ABP2,ABBCR,QBMPR,BHMS,EWAV,VAEW2014,QABBSPZ,BIP,Aguilar,Pepe,Boucaud,ACFP,DSE-BSE200912,QCDPT-DSE11,QCDPT-DSE12,QCDPT-DSE13,QCDPT-DSE14,QCDPT-DSE15,QCDPT-DSE21,QCDPT-DSE22,QCDPT-DSE23,QCDPT-DSE24,Roberts-Hadron,CR-PDA,Williams:2015EPJA,Williams:20156,Eichmann,Eichmann:2016PPNP,Eichmann:2018PRD,AW:2018CPC,MQHuber}).
DSEs can be directly derived from the Lagrangian of QCD, and thus provides a systematic and consistent way from QCD to the phenomena.
To solve the DSE for  the quark propagator, i.e., the gap equation, people need to know the information of gluon propagator and the quark-gluon vertex, which satisfy their own DSEs.
However, the DSEs for the gluon propagator and the quark-gluon vertex includes  other three- and four-point green  functions~\cite{RW,Williams:2015EPJA,Williams:20156,Eichmann:2016PPNP,Eichmann:2018PRD,CHFWW:2018,CLR,QCLR,GTL,SAH,ADFM,FA,FA1,SFK,AFLS}, therefore, truncation must be implemented for solving the DSEs.

A typical truncation for the DSEs is the rainbow approximation, that is, employing the bare quark-gluon vertex, $\gamma_{\mu}^{}$, in the DSEs. It has been applied widely (see, e.g., Refs.~\cite{AJ,MJ,RW,MR1997,MTR199903,Roberts-Hadron,QCDPT-DSE11,QCDPT-DSE12,QCDPT-DSE14,QCDPT-DSE21,CR-PDA,Eichmann,MJ,QCLR2,CMMR,GCL,VPDA,DGCL}) because of the simplicity and the systematization (people can define a self-consistent truncation with bare vertex also for the quark-quark scattering kernel in the Bethe-Salpeter equation (BSE)~\cite{RW,MR1997,MTR199903,Munczek,DSE-BSE200912,Roberts-Hadron,CR-PDA,Eichmann,BRS,SB} which is called the ladder approximation).
Such a rainbow-ladder (RL) approximation is known to be powerful for describing the ground-state vector- and isospin-nonzero-pseudoscalar-mesons~\cite{RW,MR1997,MTR199903,Roberts-Hadron,CR-PDA,Eichmann,CMMR,QCLR2,GCL,VPDA,DGCL}.
However, in the RL approximation, the coupling strength needs to be enhanced a lot to achieve the realistic results, therefore, constructing a more realistic framework that is able to meet with the physical coupling strength of QCD is necessary~\cite{BRS,BDTR,GHK,FW,BGP1,FP11,BFPR1}.
Consequently, great efforts and progress have been made to build the more realistic truncation scheme in the DSEs
themselves (see, e.g., Refs.~\cite{BCPQR,QABBSPZ,VAEW2014,ACFP,Aguilar,Pepe,Boucaud,BIP,Eichmann:2016PPNP,Eichmann:2018PRD,Williams:2015EPJA,Williams:20156,AW:2018CPC,CLR,QCLR}) and in the functional renormalization group approach (see, e.g., Refs.~\cite{MPS:2015PRD,CMPS:2018PRD,MQHuber}).
Nevertheless, efforts of including the non-Abelian property of QCD more explicitly and accomplishing the huge numerical task in solving the coupled equations are still required~\cite{Roberts-Williams:PC}.

The typical constraint  on the vertex comes from the symmetry, i.e., the Ward-Takahashi identity (WTI) ($k=p-q$) of the quark-gluon interaction,
\begin{equation}\label{WTI}
ik_{\mu}^{} \Gamma_{\mu}^{}(q,p) = S^{-1}(p) - S^{-1}(q) \, ,
\end{equation}
With the WTI, the longitudinal part of the vertex can be fixed, that is, the Ball-Chiu vertex~\cite{BC,GTL}.
Noticing that this WTI is the Abelian approximation, while the original  relation is the Slavnov-Taylor identity (STI)~\cite{MP} which reads:
\begin{eqnarray}\label{STI}
\begin{split}
ik_{\mu}^{} \Gamma_{\mu}(q,p) = & \; F^{g}(k^2)\{[1-B(p,q)]S^{-1}(p) \\
 & -S^{-1}(q)[1-B(p,q)]\} \, ,
\end{split}
\end{eqnarray}
with $F^{g}(k^2)$ the dressing function of ghost propagator and $B(p,q)$ the quark-ghost scattering kernel.

The Eq.(\ref{WTI}) or the Eq.(\ref{STI}) can only offer constraints on the 4 longitudinal components of
the quark-gluon vertex, while the full quark-gluon vertex also includes 8 transversal structures as listed in Appendix.
Through considering the anomalous magnetic moment, people introduced a phenomenological vertex
including two transversal structures besides the Ball-Chiu ansatz~\cite{CLR} which reads
\begin{equation}\label{clr1}
\Gamma_{\mu}^{}(q,p) = \Gamma^{BC}_{\mu}(q,p) + \Gamma^{ACM}_{\mu}(q,p) \, ,
\end{equation}
\begin{eqnarray}\label{clr2}
\begin{split}
& \Gamma^{BC}_{\mu}(q,p) = \Sigma_{A}^{}(q^2,p^2) L^{1}_{\mu}(q,p)
                                                            + \Delta_{A}^{}(q^2,p^2) L^{2}_{\mu} (q,p) \qquad \\
& \qquad \qquad \qquad +  \Delta_{B}^{} (q^2,p^2) L^{3}_{\mu} (q,p) \, ,  \\
& \Gamma^{ACM}_{\mu}(q,p) = b_{5}(q,p)T^{5}_{\mu}(q,p) + b_{8}(q,p) T^{8}_{\mu}(q,p) \, ,  \\
\end{split}
\end{eqnarray}
where $t=p+q$, $\Sigma_{\phi}^{}(q^2,p^2) = [ \phi(q^{2}) + \phi(p^{2}) ]/2$ and $\Delta_{\phi}(q^{2} , p^{2}) = [\phi(q^{2}) - \phi(p^{2})]/ [q^{2} - p^{2} ]$, and the tensors $L^{i}_{\mu}(q,p)$ and $T^{j}_{\mu}(q,p)$
are shown in the Appendix.
These transversal structures have been confirmed by considering the transverse Ward-Green-Takahashi identity (TWGTI)~\cite{QCLR}. This scheme with the  phenomenological vertex has been verified
to be practically equivalent to the top-down approach~\cite{BCPR} and usually referred to as the CLRQ vertex.

In this paper, we construct a scheme that combines the DSEs of the quark propagator
and the quark-gluon vertex with all the 12 Lorentz structures.
Since it has been analyzed that the non-Abelian interaction is crucial in the quark-gluon vertex~\cite{RW,BCPQR,ACFP,Aguilar,Williams:2015EPJA,Williams:20156,Eichmann:2016PPNP,AW:2018CPC,Eichmann:2018PRD,CHFWW:2018,CLR,QCLR,GTL,SAH,ADFM,FA,FA1,SFK,AFLS,FW,BAGT,AFL11,LFA11,FW:20167,BHKRT11,BCPQR11,ABP11,AP11,ACFP12,Berges,BT,WF},
we take then here the contribution of the three-gluon vertex into account in the DSEs of
the quark-gluon interaction vertex.
After solving the coupled equations, we obtain the momentum and the current quark mass dependence
of the quark propagator and the quark-gluon vertex.
Our obtained result of the quark condensate is consistent with those from other computations,
and the result for the vertex is consistent with the CLRQ vertex qualitatively.
Since this scheme is systematically derived from the QCD Lagrangian,
it is easy to be generalized to construct the quark-quark scattering kernel,
which is useful for formalizing the Bethe-Salpeter equations beyond RL approximation
and constructing a realistic scheme for computing hadron properties.

The remainder of this paper is organized as the follows.
In Section~\ref{Sec:DSEs}, we introduce briefly the coupled DSEs for the quark propagator and the quark-gluon interaction vertex.
In Section~\ref{Sec:Algorithm}, we describe the details of solving the coupled DSEs.
In Section~\ref{Sec:Solutions}, we give the obtained momentum dependence of the quark propagator and the quark-gluon
interaction vertex from the coupled DSEs at several values of the current quark mass.
In Section~\ref{Sec:BuildingNewScheme}, a parameterized expression of the quark-gluon interaction vertex is given.
A practical truncation scheme to solve the quark propagator's DSE is then built.
In Section~\ref{Sec:Applications}, the truncation method is applied to calculate the anomalous chromo- and electro-magnetic moment of the quark.
Finally a summary is given in Section~\ref{Sec:Summary}.

\section{The coupled DSEs for the quark propagator and the quark-gluon vertex}
\label{Sec:DSEs}

\subsection{The coupled DSEs}

The quark's gap equation in momentum space is usually written as
\begin{equation}\label{DS1}
S(p)^{-1} = \left(i  \slashed{p} +m_{f}^{} \right)+\Sigma(p), \\
\end{equation}
with
$$\Sigma(p) = C_{F}^{} \int^\Lambda \frac{d^4 q}{(2\pi)^4}g
\gamma_\mu S(q)G_{\mu\nu}(p-q)g\Gamma_{\nu}(p,q) \, ,  $$
where $S(p)$ is the quark propagator, $G_{\mu\nu}(p-q)$ is the gluon propagator, $\Gamma_{\nu}(p,q)$ is the quark-gluon vertex and $g$ is the running coupling of the QCD.
The constant $C_{F}^{}=\frac{N_c^2-1}{2N_c}$ is from the color structure. $m_f$ is the current quark mass.

The quark propagator can be expressed as
\begin{equation}\label{quark1}
S(p)=\frac{Z(p^{2})}{i \slashed{p} + M(p^{2})} = \frac{1}{i \slashed{p} A(p^2) + B(p^2) } \, ,
\end{equation}
where $M(p^2)$ is  the dynamical mass of the quark.

\begin{figure*}[hbt]
\centering
\includegraphics[width = 0.75\textwidth]{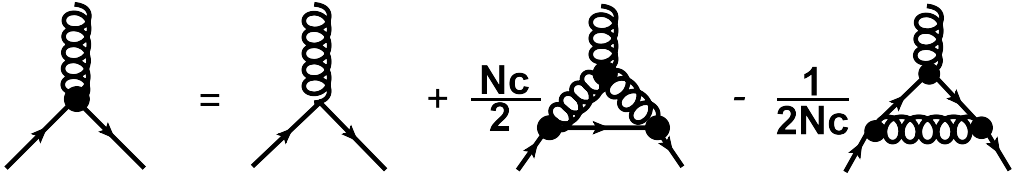}
\vspace*{-3mm}
\caption{\label{fig:vertex}The equation of motion of the quark-gluon vertex as derived from a 3PI effective action. The propagators and solid vertexes in the loops are fully dressed.}
\end{figure*}

The quark-gluon vertex satisfies its own DSE, which involves 3-point Green's functions and the full propagators,
and the full gluon propagator involves 3-point and 4-point Green's functions and quark and ghost loops~\cite{Berges}.
In order to solve such a set of coupled equations easier, one can use the equation of motion of the quark-gluon vertex as derived from the 3PI effective action~\cite{AFLS} (see Fig.~\ref{fig:vertex}),
\begin{equation}\label{DS2.1}
\Gamma_{\nu}(p,q)=\gamma_\nu+\Lambda^{A}_\nu(p,q)+\Lambda^{NA}_\nu(p,q),
\end{equation}
where
\begin{eqnarray}
\Lambda^{A}_\nu(p,q)&=&-\frac{g^2}{2N_c}\int\frac{d^4k}{(2\pi)^4}\Gamma_\beta S(q-p+k)\Gamma_\nu S(k)\nonumber\\
&& \Gamma_\alpha G_{\alpha\beta}(p-k),
\end{eqnarray}
\begin{eqnarray}
\Lambda^{NA}_\nu(p,q)&=&\frac{iN_cg^2}{2}\int\frac{d^4k}{(2\pi)^4}G_{\alpha\sigma}(p-k)\Gamma^{3g}_{\nu\sigma\tau}(p-k,k-q)\nonumber\\
&&G_{\tau\beta}(k-q)\Gamma_{\beta}S(k)\Gamma_{\alpha}.
\end{eqnarray}
The superscript ``A", ``NA" denotes the Abelian diagram (the fourth diagram in Fig.\ref{fig:vertex}),
the non-Abelian diagram (the third diagram in Fig.\ref{fig:vertex}), respectively.

\begin{figure}[hbt]
\centering
\includegraphics[width = 0.45\textwidth]{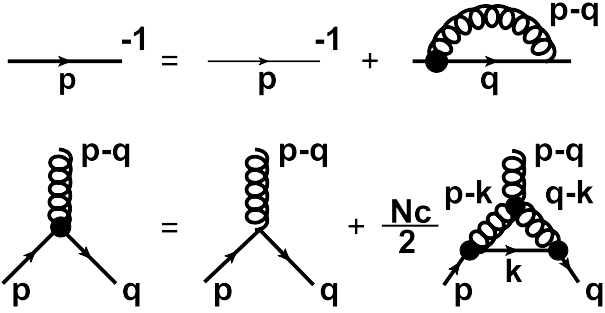}
\caption{\label{fig:feimantu}The coupled DSEs for the quark propagator and the quark-gluon vertex. The bolded lines and solid vertex in the loops are the fully dressed.}
\end{figure}

The coefficients $-\frac{1}{2N_c}$ and $\frac{N_c}{2}$ come from the color space.
It is evident that the contribution from the Abelian diagram is suppressed by a factor of $\frac{1}{N_c^2}$ compared to the non-Abelian one, thus is subleading.
Direct numerical calculations~\cite{BHKRT11,AFLS,WF,BT} have in fact also shown such a result.
Therefore, in this work, except for the bare one, we just consider the contribution from the
non-Abelian diagram that includes the three-gluon vertex.
After setting this, the coupled DSEs can be expressed diagrammatically in Fig.~\ref{fig:feimantu}.

Now, in this coupled system, the quark-gluon vertex and the quark propagator are all dressed ones with the complete 14 Lorentz structures (2 structures for quark propagator and 12 for the vertex).

\subsection{The Renormalization}

For the quark-gluon vertex, which can be expressed as
\begin{equation}\label{vertex1}
\Gamma_{\mu}^{}(q,p)= f_{1}^{} (q,p)\gamma_{\mu}^{} + \Lambda_{\mu}^{}(q,p),
\end{equation}
where $\Lambda_{\mu}^{}$ denotes all the contributions from the other tensors except the $\gamma_{\mu}^{}$, which are finite, only the first term $f_{1}^{}(q,p)$ needs to be subtracted for its UV divergence.
The renormalization condition for the quark-gluon vertex can then be chosen as
\begin{equation}
Z_{1F}^{} +f_{1}^{} (p,p)|_{p^2=\mu^2}^{} = 1 \, ,
\end{equation}
and the renormalized quark-gluon vertex reads,
\begin{equation}
\Gamma^{r}_{\mu}(q,p) = \big{(} f_{1}^{} (q,p) + Z_{1F}^{} \big{)} \gamma_{\mu}^{} + \Lambda_{\mu}^{} (q,p) \, ,
\end{equation}

For the quark propagator, the renormalized gap equation reads,
\begin{equation}
S^{r}(p)^{-1}=Z_{2}^{} \big{(} i \slashed{p} + Z_{4}^{} m_{f}^{} \big{)} + Z_{1}^{} \Sigma(p) \, .
\end{equation}
If expressing the self energy $\Sigma(p) = i \slashed{p}\Sigma_{1}^{}(p^{2}) + \Sigma_{2}^{}(p^{2})$ with
$\Sigma_{1}^{}(p^{2})$ and $\Sigma_{2}^{}(p^2)$ the scalar functions,
the solution of the gap equation can be expressed as,
\begin{eqnarray}\label{AB}
\begin{split}
A(p^{2})  = & \; Z_{2}^{} + Z_{1}^{} \Sigma_{1}^{}(p^{2}) \, , \\
B(p^{2})  = & \; Z_{2}^{}Z_{4}^{} m_{f}^{} + Z_{1}^{} \Sigma_{2}^{}(p^{2}) \, .
\end{split}
\end{eqnarray}
Based on considering the WTI, we set $Z_{1}^{} = Z_{2}^{}$.
The renormalization condition for the quark propagator is then,
\begin{equation}
S^{r}(\mu)^{-1} = i \slashed{\mu} + m_{f}^{} \, ,
\end{equation}
and the renormalization factor reads
\begin{equation}
Z_{2}^{} = \frac{1}{\Sigma_{1}^{}(\mu^{2}) +1}, \; \;
Z_{2}^{} Z_{m}^{} m_{f}^{} = m_{f}^{} - \frac{\Sigma_{2}^{} (\mu^{2})}{\Sigma_{1}^{}(\mu^{2}) + 1} \, .
\end{equation}

\section{Solving the coupled DSEs}
\label{Sec:Algorithm}

\subsection{Formulation  of the quark-gluon vertex}

The Lorentz structure of the quark-gluon vertex in Eq.(\ref{DS1}), $\Gamma_{\nu}^{}(p,q)$,
should be expressed as a combination of all the 12 independent Lorenz structures.
For convenience, we choose the tensor structures which have been used in Refs.~\cite{QCLR,CLR,BCPQR} as:
\begin{equation}
\Gamma_{\mu}^{}(q,p) = \Gamma^{L}_{\mu}(q,p) + \Gamma^{T}_{\mu}(q,p) \, ,
\end{equation}
with
\begin{eqnarray}\label{tensors1}
\begin{split}
\Gamma^{L}_{\mu}(q,p) & = & \sum^{4}_{i=1} a_{i}^{} (q,p) L^{i}_{\mu}(p+q,p-q), \\
\Gamma^{T}_{\mu}(q,p) & = & \sum^{8}_{i=1} b_{i}(q,p) T^{i}_{\mu}(p+q,p-q) \, .
\end{split}
\end{eqnarray}
where the matrix-valued tensors $L^{i}_{\mu}(q,p)$ and $T^{i}_{\mu}(q,p)$ are given, respectively,
as Eq.(\ref{tensors11}) and Eq.(\ref{tensors12}) in Appendix.

This is the standard basis that includes the Ball-Chiu ansatz, which satisfies the WTIs,
and the 8 transversal components hold relation $k_{\mu}^{} \Gamma^{T}_{\mu}(q,p)\equiv 0$.
However, this  basis are not completely orthogonal, which brings in  the difficulty for numerical calculation.  Therefore, we also implement here another set of basis which are given in Eq.(\ref{tensors21}) and Eq.(\ref{tensors22}) in the Appendix.
Therefore, the quark-gluon interaction vertex can be expressed as
\begin{equation}
\Gamma_{\mu}^{}(q,p) = \Gamma'^{L}_{\mu}(q,p) + \Gamma'^{T}_{\mu}(q,p) \, ,
\end{equation}
with
\begin{eqnarray}\label{tensors2}
\begin{split}
\Gamma'^{L}_{\mu}(q,p) & = & \sum^{4}_{i=1}a'_{i} (q,p) L'^{i}_{\mu}(p+q,p-q) \, , \\
\Gamma'^{T}_{\mu}(q,p) & = & \sum^{8}_{i=1}b'_{i}(q,p) T'^{i}_{\mu}(p+q,p-q) \, .
\end{split}
\end{eqnarray}
The coefficients of the orthogonalized basis can be easily obtained via the following projection:
\begin{eqnarray}\label{DS2}
\begin{split}
a'_{i}(q,p)  & = \! \! & \! \! \frac{L'^{i}_{\mu}(p+q,p-q)\cdot\Gamma_{\mu}(q,p)}{L'^{i}_{\mu}(p+q,p-q)\cdot L'^{i}_{\mu}(p+q,p-q)} \, , \qquad \\
b'_{i}(q,p)  & = \! \! & \! \! \frac{T'^{i}_{\mu}(p+q,p-q)\cdot\Gamma_{\mu}(q,p)}{T'^{i}_{\mu}(p+q,p-q)\cdot T'^{i}_{\mu}(p+q,p-q)} \, . \qquad
\end{split}
\end{eqnarray}
After solving the equations, we can convert these coefficients into those with the standard basis
for comparison with other results. The relation of the two sets can be given explicitly as
\begin{eqnarray}\label{11}
\begin{split}
& a_{1}^{} = a'_{1} - \frac{(t\cdot k)^{2}}{2k^{2}} a'_{2} \, , \qquad & a_{2}^{} = a'_{2} \, , \qquad \\
& a_{3}^{} = a'_{3} \, , \qquad\qquad\qquad\,\,\, & a_{4}^{} = a'_{4} \, ; \qquad
\end{split}
\end{eqnarray}
and
\begin{eqnarray}\label{22}
\begin{split}
b_{1}^{}  = & \, \frac{2}{k^{2} }( a'_{3} - b'_{5} ) \, ,  \\
b_{2}^{}  = & \, \frac{1}{k^{2} }(-a'_{2} + 6 b'_{2}) \, ,  \\
b_{3}^{}  = & \, \frac{1}{k^{2}} \Big{[} -a'_{1} + \frac{(t\cdot k)^{2}}{2k^{2}}a'_{2} + b'_{1}  \\
& - \Big{(} t^{2} + \frac{2(t\cdot k)^{2}}{k^{2}} \Big{)} b'_{2} + (t\cdot k)^{2} b'_{2} \Big{]} \, ,  \\[1mm]
b_{4}^{}  = & \, \frac{4}{k^{2}} \Big{[} -\frac{1}{t\cdot k}a'_{4} + 2b'_{6} + 2b'_{7} - b'_{8} \Big{]} \, ,  \\[1mm]
b_{5}^{}  = & \, -2\frac{(t\cdot k)^{2}}{k^{2}}b'_{6} + 2 \Big{(} t^{2} - \frac{(t\cdot k)^{2}}{k^{2}} \Big{)} b'_{7} \, ,  \\[1mm]
b_{6}^{}  = & \; t\cdot k \Big{(} \frac{3}{k^{2}}b'_{2} - b'_{3} \Big{)}  \, ,  \\[1mm]
b_{7}^{}  = & \, -\frac{2}{t\cdot k}a'_{4} + 4 b'_{6} \, ,  \\
b_{8}^{}  = & \; b'_{4} \, ,
\end{split}
\end{eqnarray}
with $t=p+q$, $k=p-q$.

\subsection{Necessary Input}

To solve the coupled DSEs, we need  the input of the gluon propagator, the running coupling and the three-gluon vertex. In this work, we use directly a set of results from the ab initio computations~\cite{Cucchieri,BIPS,Oliveira,ABP1,AP,ABP2,ABBCR,HG,AG,QBMPR,BHMS,QABBSPZ,EWAV}
as the input,  without adding any other parameters.

\subsubsection{The gluon propagator}

Recently, the infrared (IF) behavior of the gluon propagator has been investigated a lot, through lattice QCD simulations~\cite{Cucchieri,BIPS,Oliveira} and DSE approach~\cite{ABP1,AP,ABP2,ABBCR}.
All the computations have reached a common idea
that  a mass scale is dynamically generated in the IF region of the gluon propagator.
In this article, we choose the numerical result  and the fitted formula given in Ref.~\cite{AP}
\begin{equation}\label{gluon1}
G_{\mu\nu}^{}(k)= \Big{(} \delta_{\mu\nu}-\frac{k_\mu k_\nu}{k^2} \Big{)} \Delta(k^2),
\end{equation}
where
\begin{equation}\label{gluon2}
\Delta^{-1}(k^2)=m^2(k^2)
+k^{2}\Big{[} 1+\frac{13C_{A}^{}g_{1}^{2}}{93\pi^{2}} \ln\Big{(} \frac{k^{2} + \rho_{1}^{} m^{2}(k^2)}{\mu^2} \Big{)} \Big{]} \, ,
\end{equation}
the $m^{2}(k^2) = \frac{m^{4}}{k^{2} + \rho_{2}^{} m^{2}}$ is the dynamical mass of the gluon,
$\mu=4.3\,$GeV is the renormalization point which is generally taken in  lattice QCD simulations
for gluon. The fitted  parameters are
\begin{equation}\label{gluon3}
m=500\,\textrm{MeV}\, ,\;\;  g_{1}^{2}=5.68, \;\; \rho_{1}^{} =60\, , \;\; \rho_{2}^{} = 1.6 \, .
\end{equation}
Consequently, the gluon obtains a mass scale at zero momentum, $m(0)=395\,$MeV, and goes to the asymptotic form in ultraviolet (UV) region.

\subsubsection{The running coupling}

It has been shown through the computation of both functional renormalization group approach~\cite{HG} and  DSE approach~\cite{AG,QBMPR}
that the running coupling  $\alpha(k^2)=g^2(k^2)/4\pi$ is a constant in the IF region,
and can be expressed explicitly as~\cite{QBMPR}:
\begin{equation}\label{coupling1}
\alpha(k^{2})=\frac{1 - d_{0}^{} k^{2} \ln\frac{k^{2}}{\Lambda^{2}_{T}} + d_{1}^{} k^{2} + d_{2}^{} k^{4}}{1 + b_{1}^{} k^{2} + c_{0} d_{2}^{} k^{4} \ln\frac{k^{2}}{\Lambda^{2}_{T}}} \pi \, ,
\end{equation}
where
\begin{eqnarray}
\begin{split}
& d_{0}^{} = \frac{14.4}{8\pi} \, \textrm{GeV}^{-2}, \quad & d_{1}^{} =3.56 \, \textrm{GeV}^{-2} , \\
& d_{2}^{} = 2.85\, \textrm{GeV}^{-4}, \quad & b_{1}^{} =7.25\, \textrm{GeV}^{-2} ,\\
& c_{0}^{} = \frac{11-\frac{2N_f}{3}}{4\pi}.
\end{split}
\end{eqnarray}
After considering the running coupling's behavior in UV region~\cite{AG},
one can set the $\Lambda_{T}^{}$ as $\Lambda_{T}^{} = 130\,$MeV.

\subsubsection{The three-gluon vertex}

There have been a lot of works to investigate the behavior of the three-gluon vertex through different
methods~\cite{BHMS,QABBSPZ,EWAV,VAEW2014} and shown that it involves a ``zero-crossing" in the IF region.
Therefore, when solving the DSEs of the quark propagator and the quark-gluon interaction vertex,
we choose the following form as the input of the three-gluon vertex
\begin{eqnarray}
\Gamma^{3g}_{\nu\sigma\tau}& & (p-k,k-q) = G(l^2) \big{[} (q+k-2p)_{\tau} \delta_{\nu\sigma} \qquad \qquad \nonumber\\ && \qquad +(p+q-2k)_{\nu} \delta_{\sigma\tau}+(k+p-2q)_{\sigma} \delta_{\nu\tau} \big{]} \, ,
\end{eqnarray}
where
$$ G(l^2) = \left\{ \begin{array}{ll}
0.865(\log_{10}(l^2/2+1))^{\frac{17}{44}} & \;\; l^{2}<55.47\,\textrm{GeV}^2 \, , \\[1mm]
1 & \;\; l^{2} > 55.47\,\textrm{GeV}^{2} \, ,  \end{array}  \right. $$
in which $l^2=(p-q)^2+(k-p)^2+(k-q)^2$. This expression guarantees that $G(l^2)=0$ when $l^2=0$. As $l^2$ increases, $G(l^2)$ becomes larger and goes to 1 when $l^{2} > 3\times 4.3^{2}\, \textrm{GeV}^{2}$.

\subsection{Approximation on the dependence of the angle}

Previous works have shown that the coefficients of the Lorentz structures of the quark-gluon vertex do not have considerable dependence on the  angle between the two momentums~\cite{BC,CLR}.
We then intend to employ the assumption:
\begin{eqnarray}\label{angle}
\begin{split}
a_{i}^{} (q^2,p^2) \! & \! = \! & \! a_{i}^{} (q^{2}, p^{2}, \bar{\theta}) \, , \\
b_{j}^{} (q^2,p^2) \! & \! = \! & \! b_{j}^{} (q^{2}, p^{2}, \bar{\theta}) \, .
\end{split}
\end{eqnarray}
with $\bar{\theta}$ being a fixed value for the angle.

\begin{figure}[hbt]
\begin{center}
\includegraphics[width=0.43\textwidth]{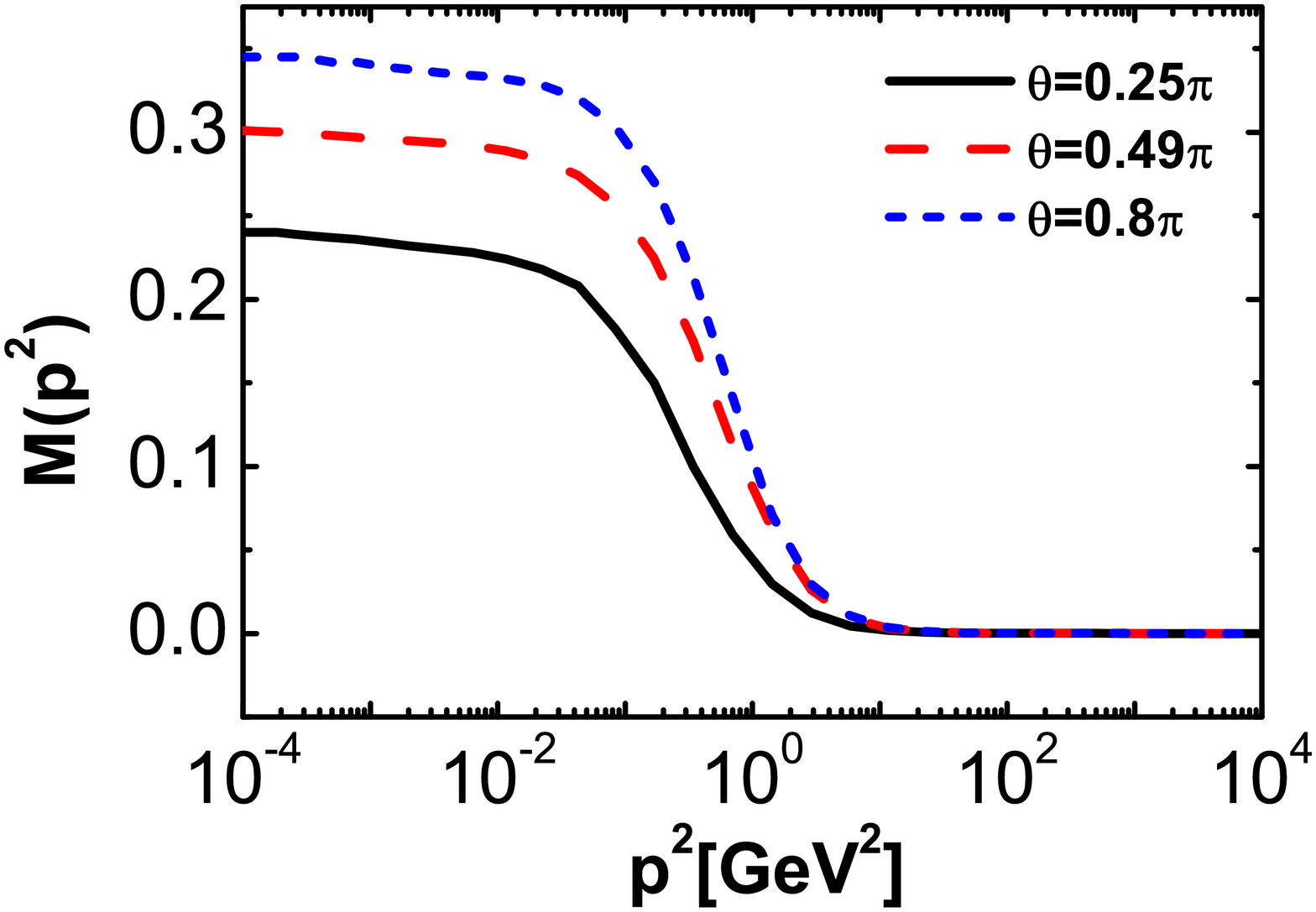}
\includegraphics[width=0.43\textwidth]{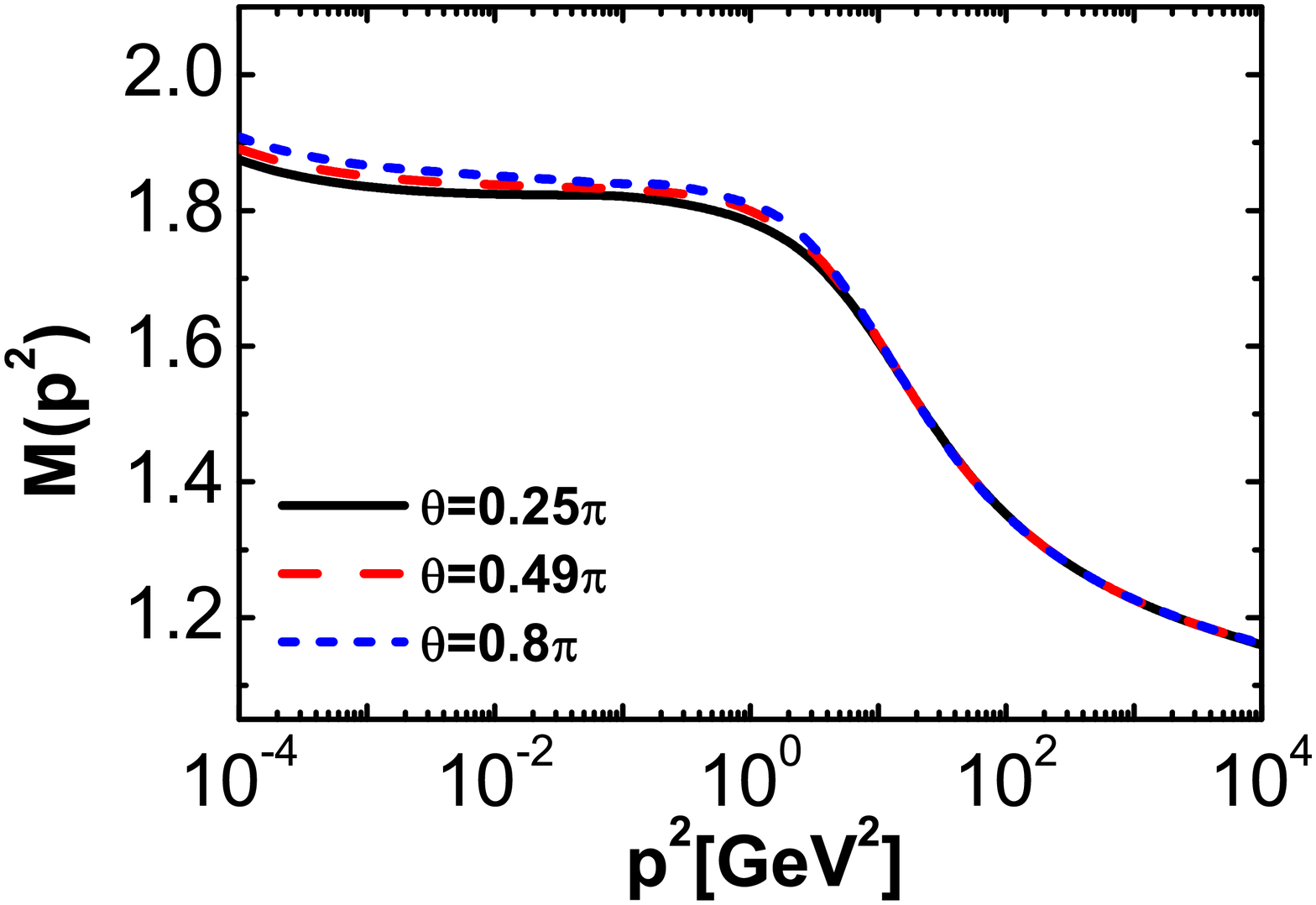}
\end{center}
\vspace*{-6mm}
\caption{Calculated dynamical masses $M(p^2)$ (in unit GeV) in case of $\theta=0.25\pi$, $\theta=0.49\pi$ and
$\theta=0.8\pi$.  \emph{Upper panel}: the result for $m_{f}^{}=0.0\,$GeV.
\emph{Lower panel}: the result for $m_{f}^{} = 1.27\,$GeV. }
\label{fig:theta}
\end{figure}

Nevertheless, it has also been shown numerically that there exists few difference among
the quark-gluon vertex's longitudinal behaviors in different angles~\cite{ACFP}.
To be more careful, we check it by analyzing the solutions of the coupled DSEs.
The obtained results of the dynamical masses in case of $\bar{\theta}=0.25\pi$, $0.49\pi$ and $0.8\pi$
are shown in Fig.\ref{fig:theta}.
It can be found easily that, for light quark,
there exists observable difference between the dynamical masses from the coupled DSEs with different angles,
and the difference diminishes gradually as the current quark mass increases.
Then, considering the monotonous behavior of the dynamical mass $M(p^2)$ with respect to the angle,
one can still take the assumption in Eq.(\ref{angle}) with choosing the central value at $\bar{\theta}=0.49\pi$
even though the effect of the angle in case of light quark can not be completely ignored.

\section{Solutions of the coupled DSEs}
\label{Sec:Solutions}

In this section, we display and discuss the numerical results of the  quark propagator and the quark-gluon interaction vertex.
Firstly, we search the quark propagator's multiple solutions to build the complete picture of the DCSB, and analyze the current quark mass dependence to show the DCSB effect in case of that the explicit chiral symmetry breaking (ECSB) has already been there. Secondly, we calculate the quark-gluon vertex and analyze the results in some different cases.

\subsection{Quark Propagator} \label{Subsec:Solutions-QP}

\begin{figure}[hbt]
\begin{center}
\includegraphics[width=0.43\textwidth]{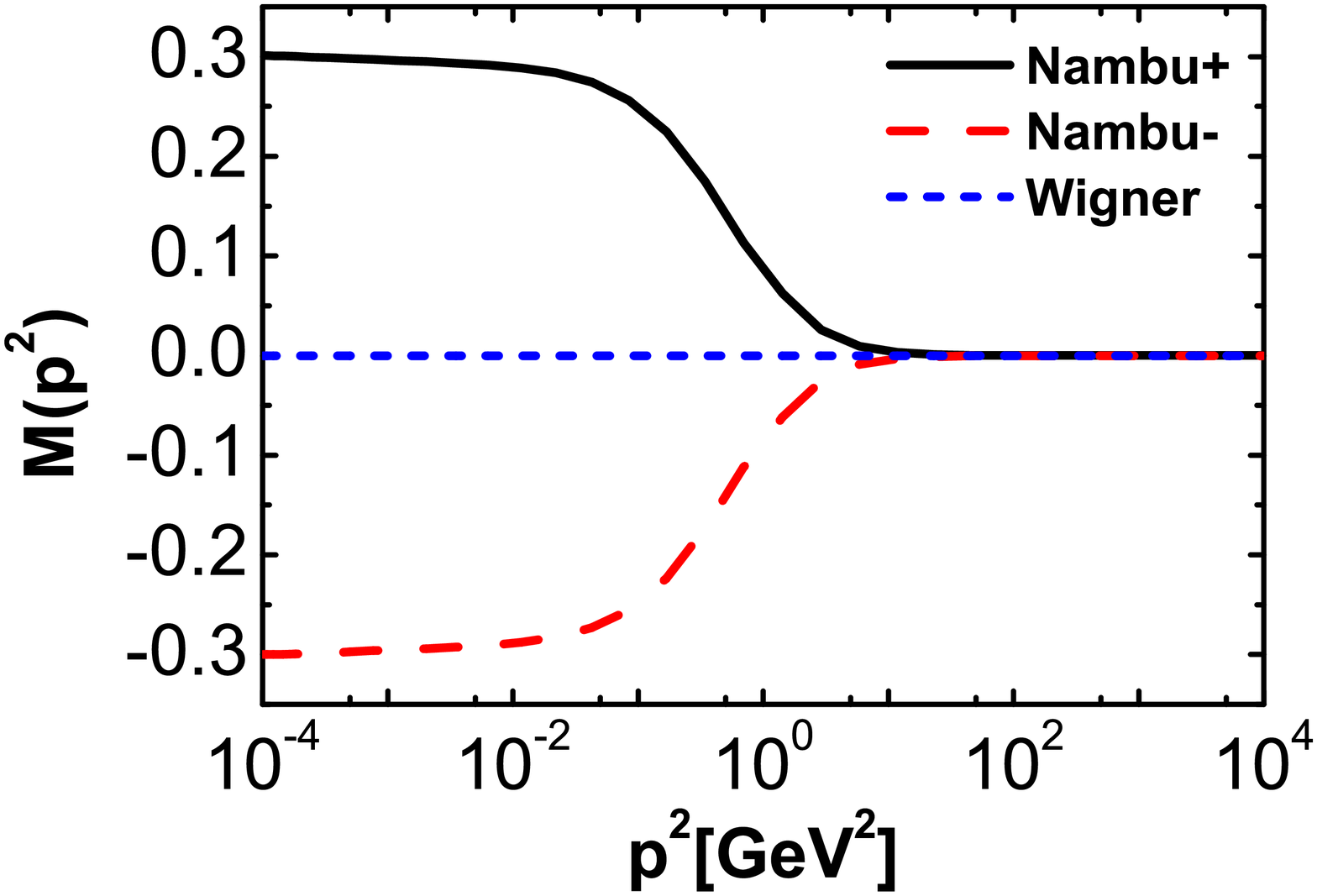}
\includegraphics[width=0.43\textwidth]{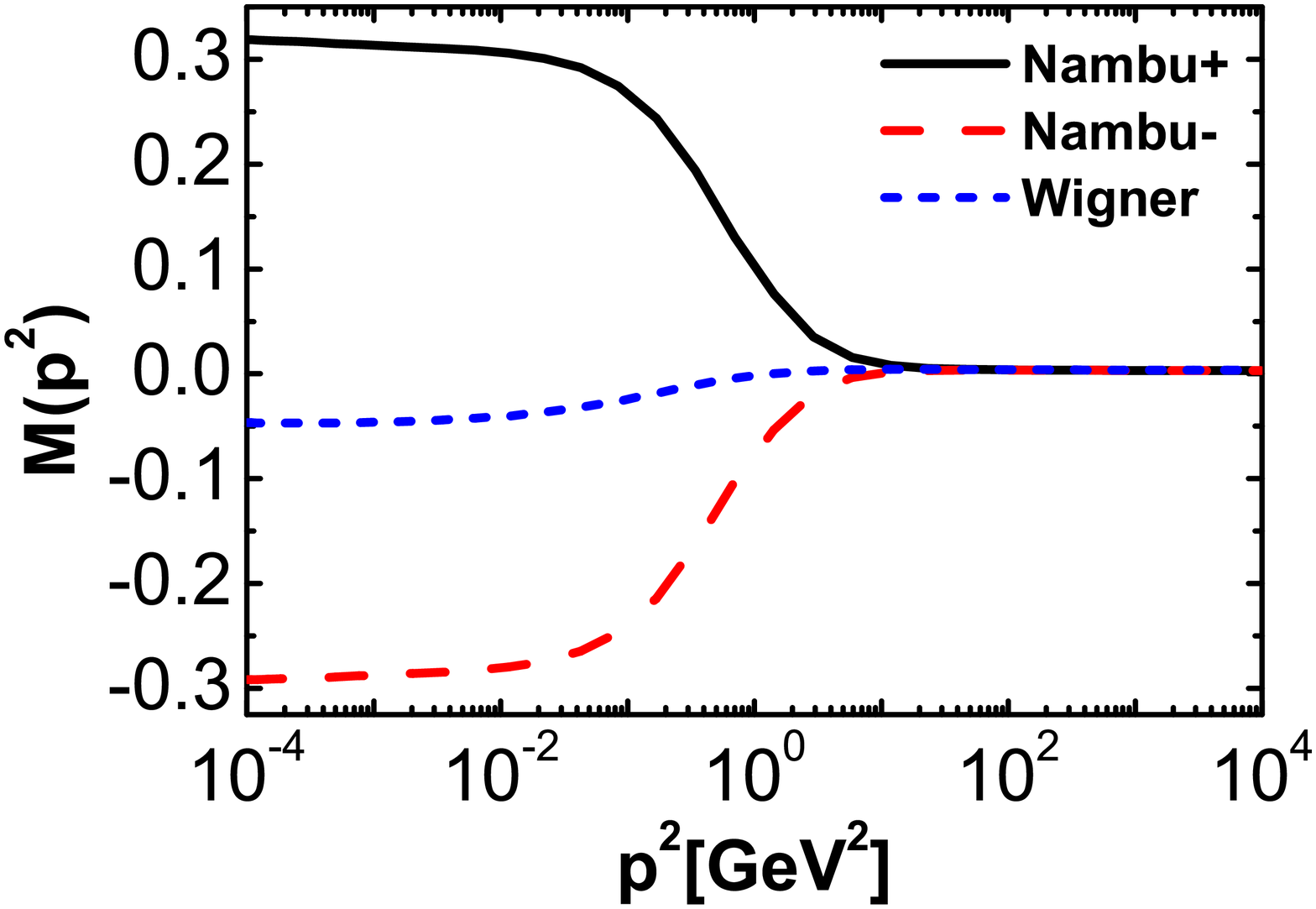}
\end{center}
\vspace*{-5mm}
\caption{Calculated multiple solutions of the dynamical mass (in unit GeV) in the quark propagator in case of chiral limit $m_{f}^{}= 0.0\,$MeV
(\emph{upper panel}), and those of a small current mass $m_{f}^{}=3.4\,$MeV (\emph{lower panel}) .}
\label{fig:multiple}
\end{figure}

It is generally believed that there are multiple solutions for the quark propagator~\cite{QCDPT-DSE11,QCDPT-DSE12,QCDPT-DSE13,QCDPT-DSE21,JGSZ,CSXJZ,XCSZ},
at least, the Nambu solutions and the Wigner solutions.
After managing the initial conditions carefully,
we obtain the multiple solutions of the quark propagator in chiral limit and also in case of a small current quark mass. The obtained results are shown in Fig.~\ref{fig:multiple}.

Fig.~\ref{fig:multiple} shows that,
in chiral limit, the dynamical mass is generated and demonstrated with the Nambu solution and its negative counterpart.
Meanwhile, there is also a meta-stable Wigner solution with vanishing mass function,
which is the chiral symmetry preserving state.
It indicates apparently that there exists DCSB.
We then compute  the quark condensate from the Nambu+ solution in the chiral limit,
and obtain  $- \langle \bar{q}q \rangle^{1/3}(m_{f}^{} = 0) = 290\,$MeV.
This is definitely consistent with the other calculations (e.g., given in Ref.\cite{FW})).
It has also been shown that, if the current quark mass is not large enough,
the two kind solutions for the quark propagator still exist~\cite{QCDPT-DSE11,QCDPT-DSE13,QCDPT-DSE21}.
The difference from that in the chiral limit is mainly that the Wigner solution in IF region becomes non-zero,
which shows the ECSB effect brought by the finite current quark mass.
And the DCSB still exists and exhibited also by the Nambu solution.
We have, in turn, $- \langle \bar{q}q \rangle^{1/3} (m_{f}^{} \! = \! 3.4\,\textrm{MeV}) \! = \! 290\,$MeV.
This is evidently consistent with that constrained by the Gell-Mann--Oakes--Renner (GOR) relation~\cite{GOR}.
Noticing that here we take directly the gluon propagator and the coupling strength from
the ab initio computation and do not include any other parameters,
the consistency indicates that our presently obtained quark-gluon interaction vertex with only the three-gluon interaction being involved would be sufficient to describe the DCSB of QCD.
%
%

\begin{figure}[hbt]
\begin{center}
\includegraphics[width=0.43\textwidth]{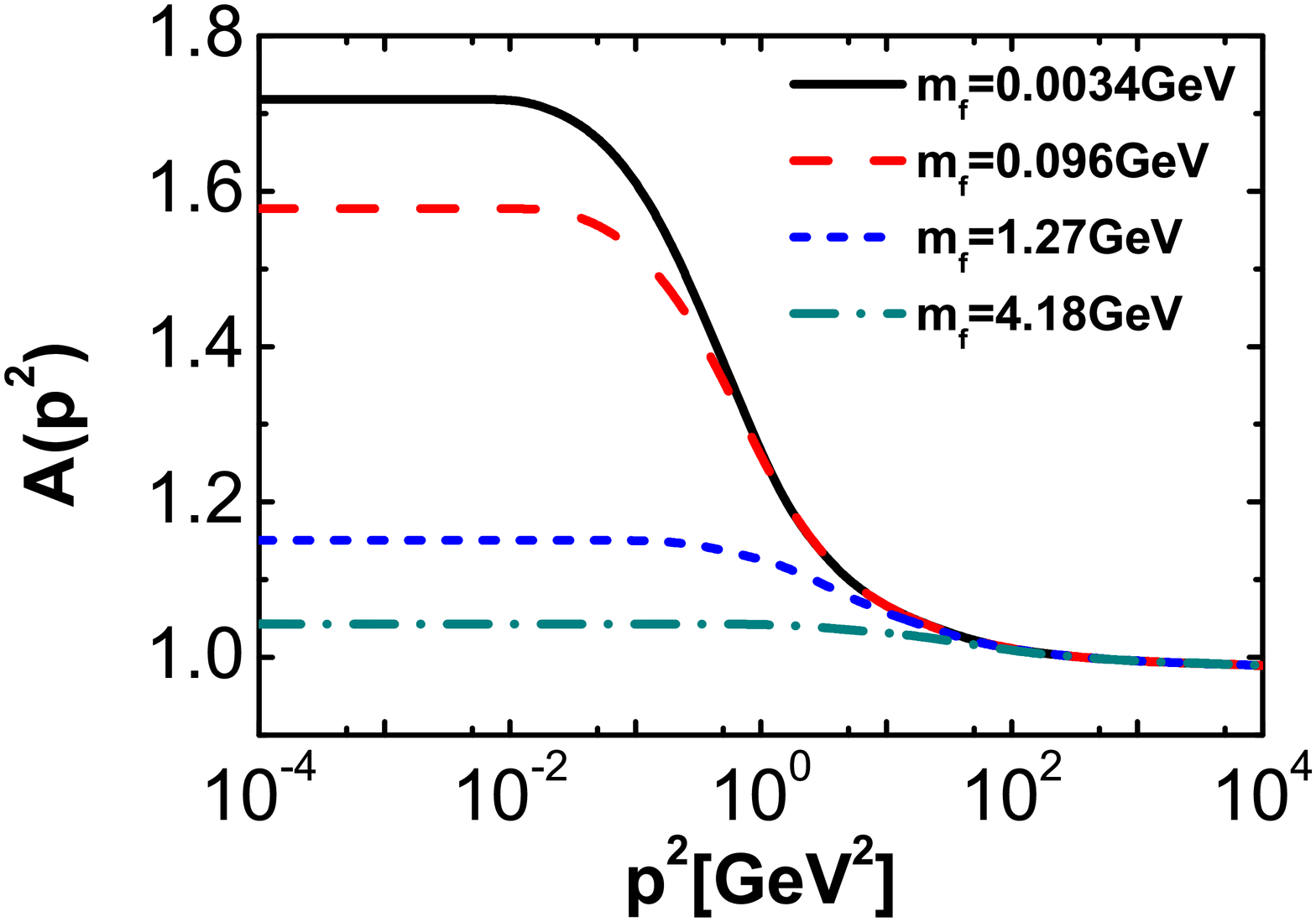}
\includegraphics[width=0.43\textwidth]{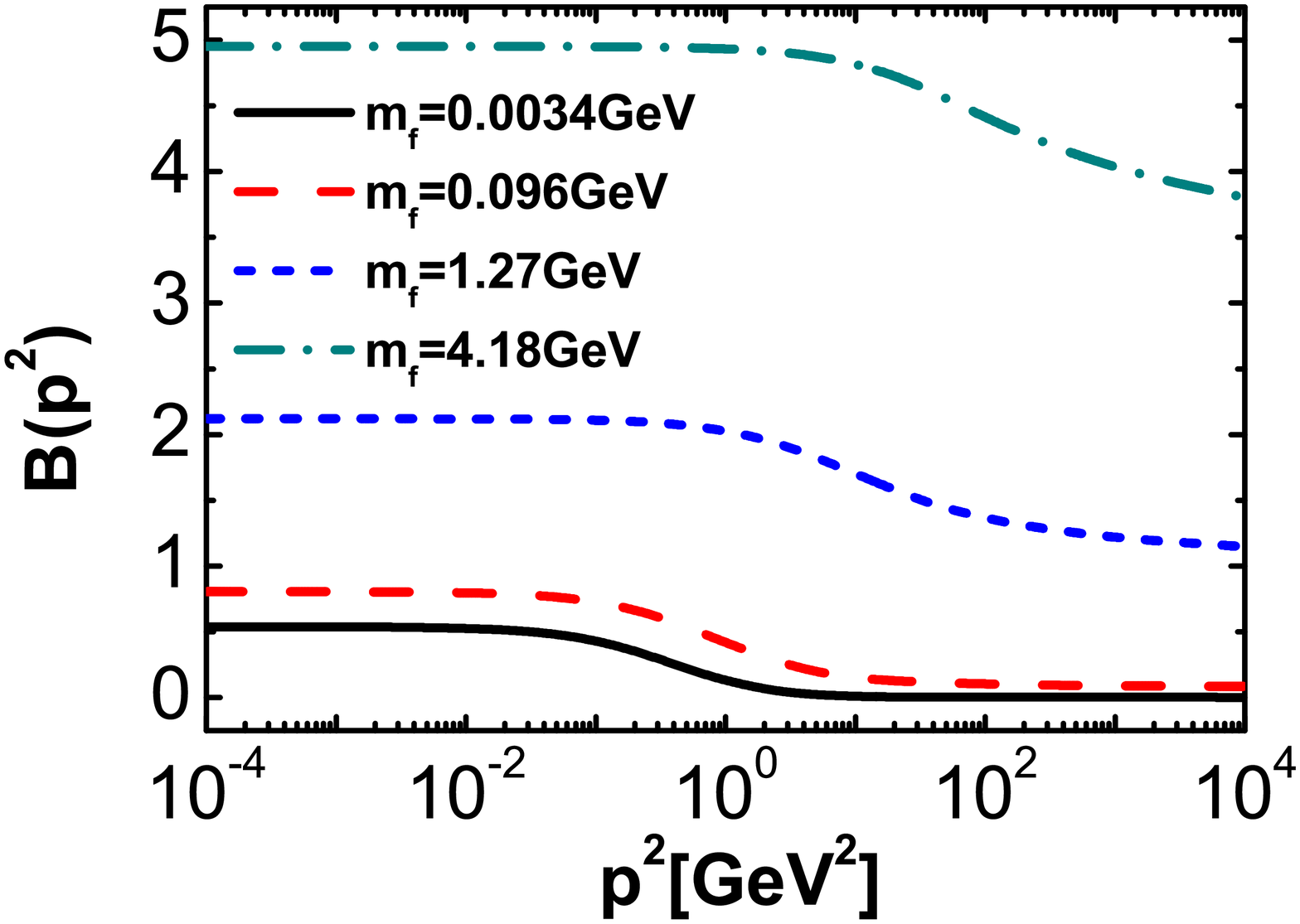}
\end{center}
\vspace*{-6mm}
\caption{ Calculated scalar functions $A(p^2)$ and $B(p^2)$ in the quark propagators
at several values of the current quark mass.}
\label{fig:diffmass}
\end{figure}

\begin{figure}[hbt]
\begin{center}
\includegraphics[width=0.43\textwidth]{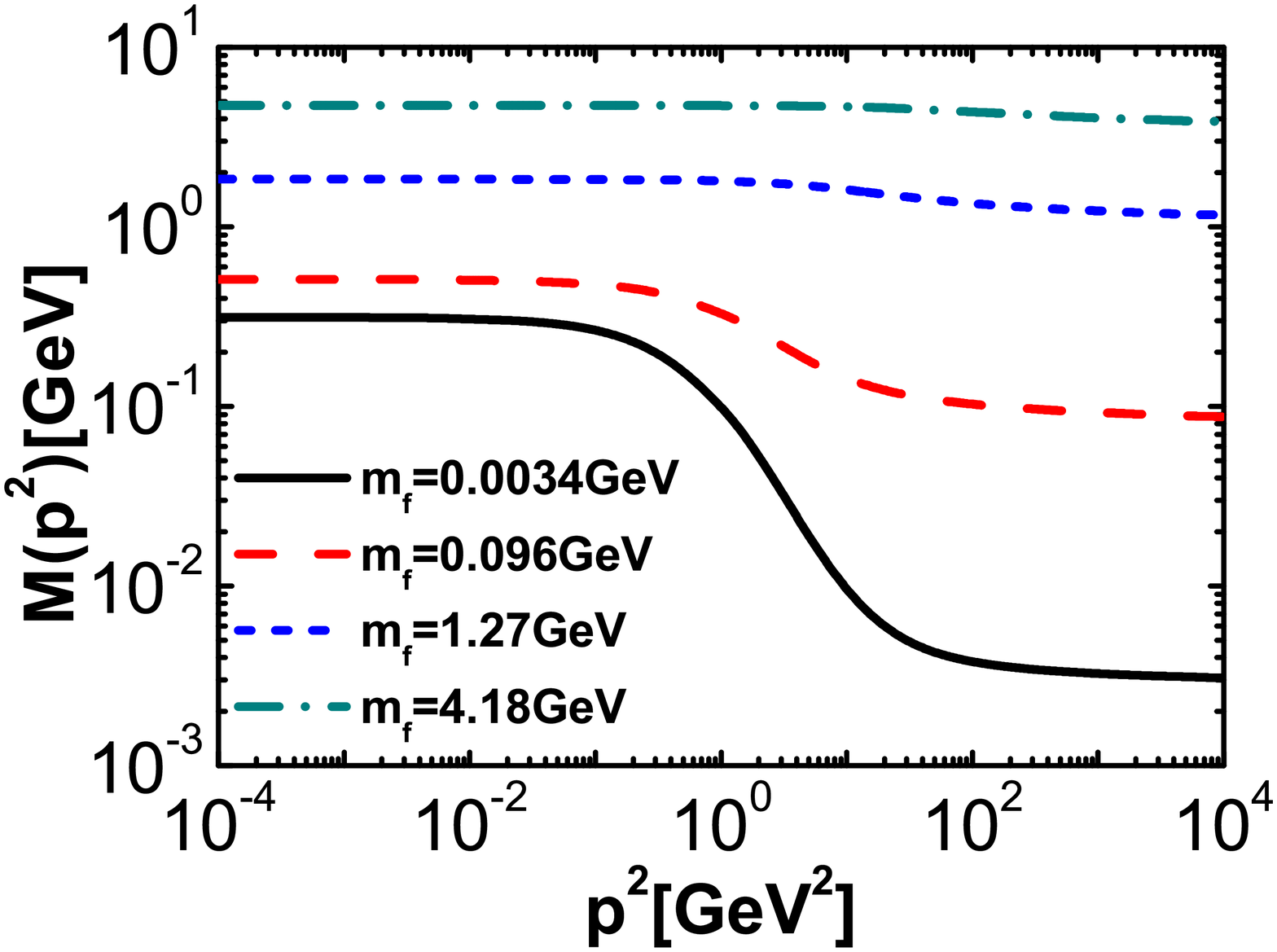}
\includegraphics[width=0.43\textwidth]{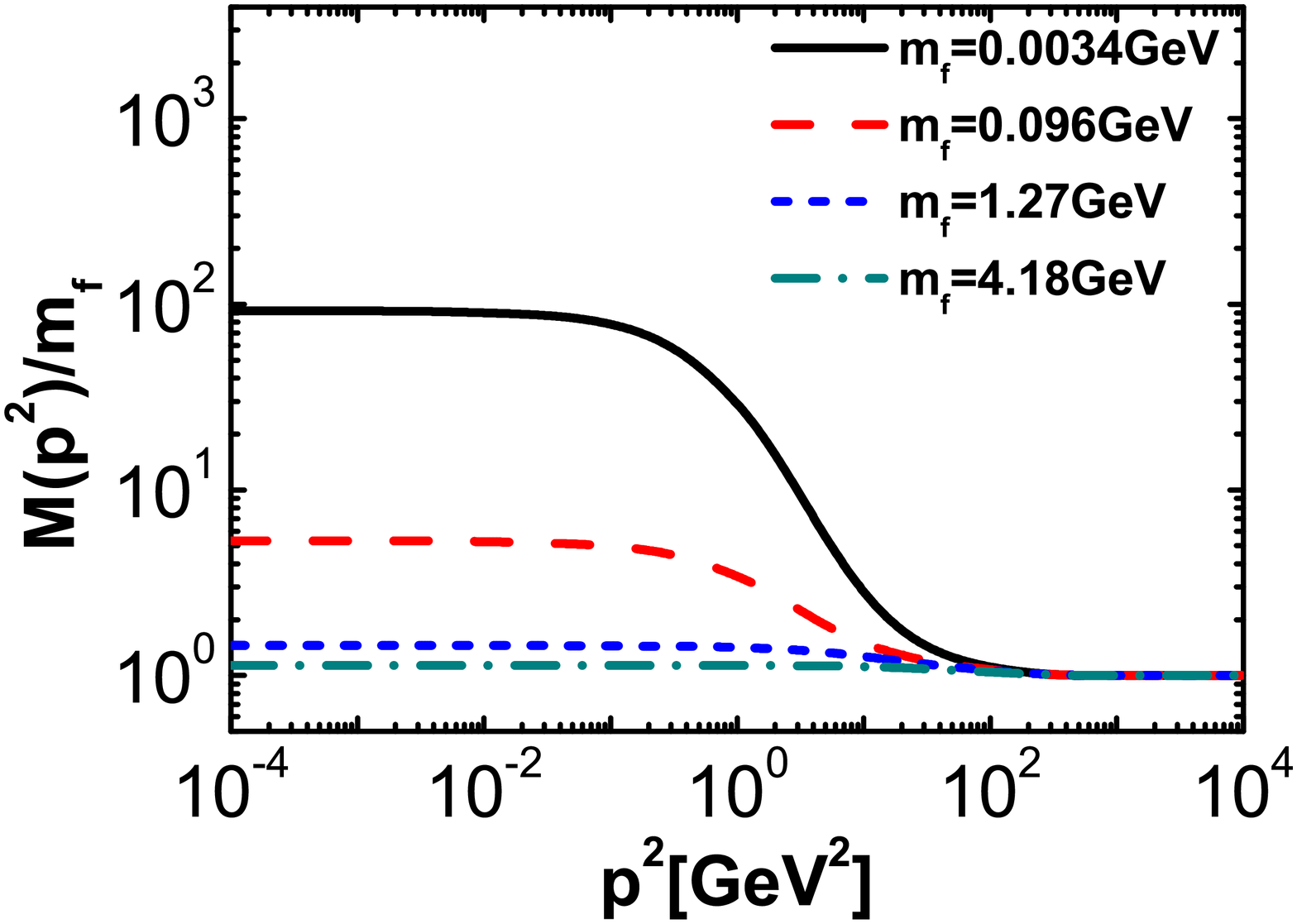}
\end{center}
\vspace*{-6mm}
\caption{ Calculated dynamical mass of the quark with several values of the current mass ({\it upper panel})
and the ratios of the dynamical mass at zero momentum with the current mass ({\it lower panel}).}
\label{fig:DynamicalMass}
\end{figure}

To show the properties of the quark propagator more thoroughly,
we also computed the solutions for different current quark masses up to that of bottom quark.
The obtained results of the scalar functions $A(p^2)$ and $B(p^2)$ in the quark propagators
at several values of the current quark mass are displayed in Fig.~\ref{fig:diffmass}.
It is apparent that, as the current quark mass $m_{q}^{}$ increases, the function $A(p^2)$ decreases monotonously  and tends to be 1 in heavy quark limit, while the mass function $B(p^{2})$ increases.
In turn, as shown in Fig.~\ref{fig:DynamicalMass},
the dynamical mass $M(p^{2}) = B(p^{2})/A(p^{2})$ increases as the current mass ascends.
However, the ratio of $M(0)/m_{q}^{}$ decreases with respect to the increasing of the $m_{q}^{}$.
This indicates that the effect of the DCSB becomes smaller as the current mass increases,
and  the ECSB effect related to the current quark mass becomes dominant
as the current mass gets as large as that of bottom quark and further.

\subsection{Quark-Gluon Interaction Vertex}   \label{Subsec:Solutions-qg-Vetex}

\subsubsection{3D distribution of the coefficients in chiral limit}

\begin{figure}[hbt]
\centerline{\includegraphics[width=0.43\textwidth]{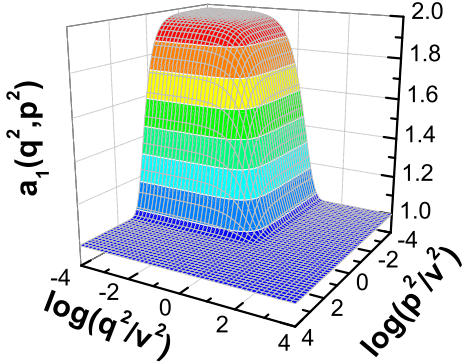}}
\vspace*{-3mm}
\caption{Calculated momentum dependence of the coefficient $a_{1}^{}$ (with the scale parameter $\textrm{v}=1.0\,$GeV).}
\label{fig:l1}
\end{figure}

\begin{figure}[hbt]
\begin{center}
\includegraphics[width=0.43\textwidth]{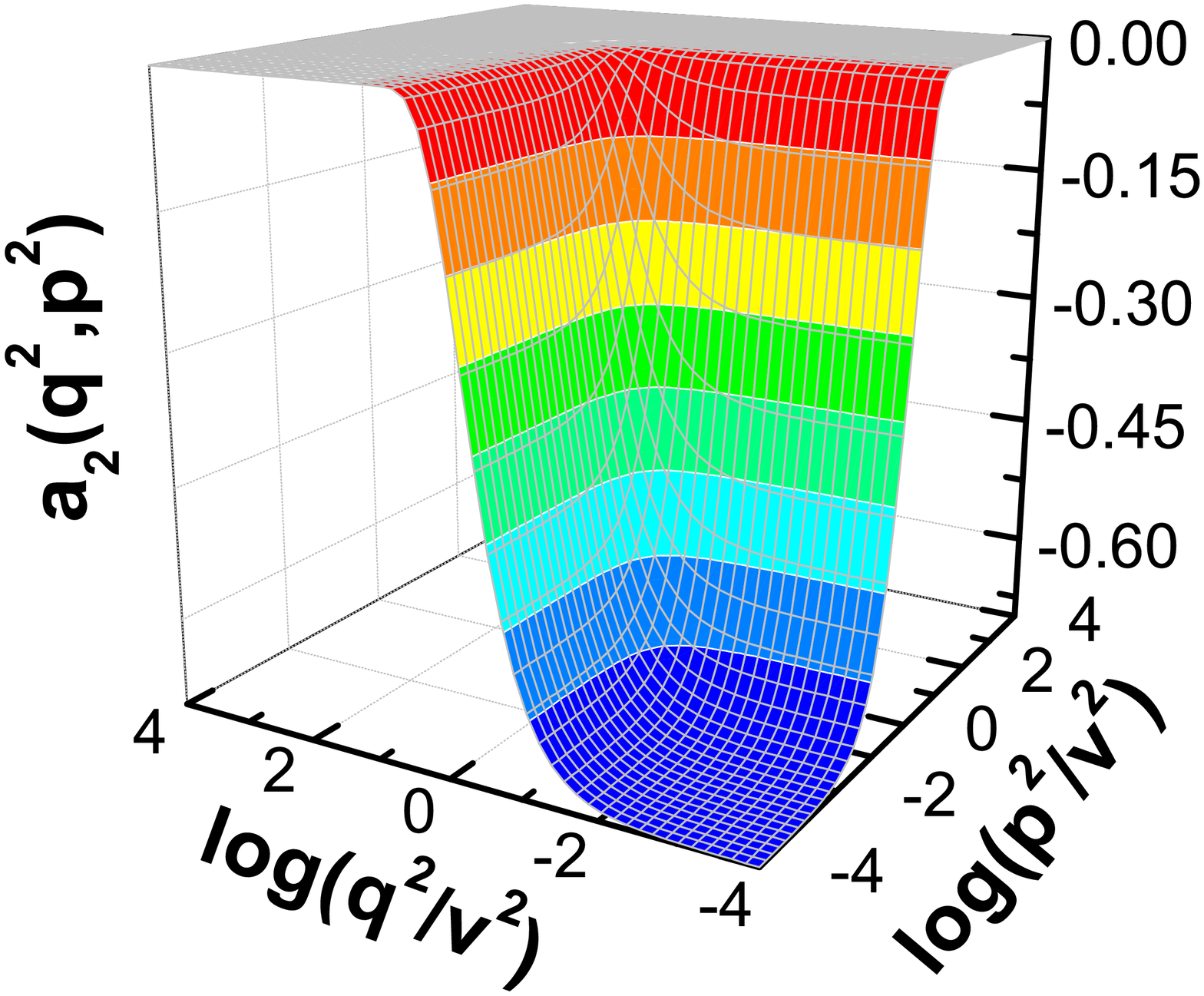}
\includegraphics[width=0.43\textwidth]{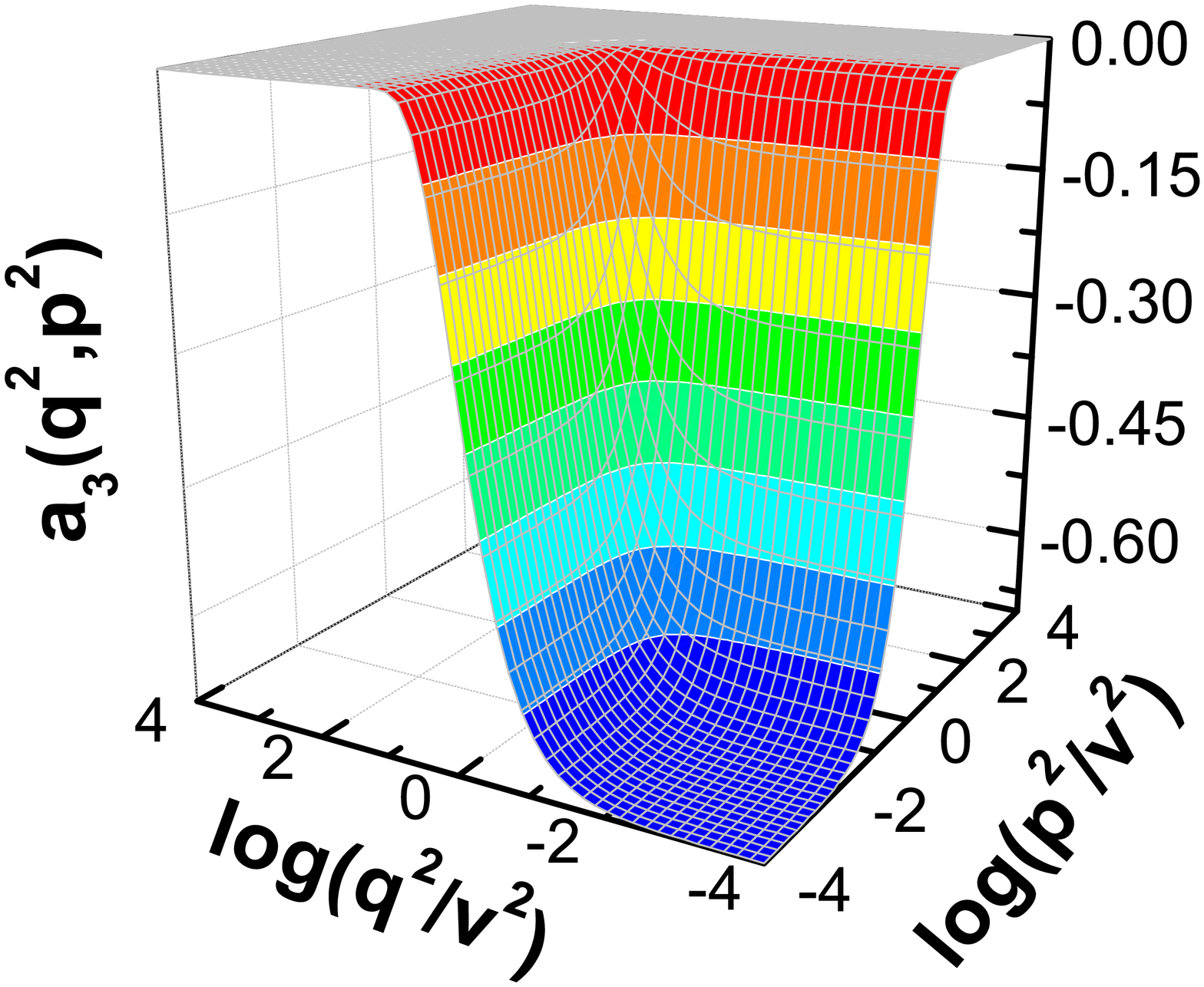}
\end{center}
\vspace*{-4mm}\caption{ Calculated momentum dependence of the coefficient $a_{2}^{}$ (\emph{upper panel})
and that of coefficient $a_{3}^{}$ (\emph{lower panel}).
The scale parameter is also taken as $\textrm{v}=1.0\,$GeV. }
\label{fig:l2l3}
\end{figure}

The obtained momentum dependence of the coefficients of the longitudinal Lorentz structures, $a_{1}^{}$, $a_{2}^{}$ and $a_{3}^{}$  are illustrated in Fig.~\ref{fig:l1} and Fig.~\ref{fig:l2l3}. It is apparent that, at arbitrary momentums
$q^{2}$ and $p^{2}$, the $a_{1}^{}$ takes always non-negative value, but the $a_{2}^{}$ and $a_{3}^{}$ get definitely non-positive values. All the amplitudes (absolute values) of the coefficients decrease monotonously
as the momentums  increase, and tend to 0 in the UV region.

\begin{figure}[hbt]
\centerline{\includegraphics[width=0.43\textwidth]{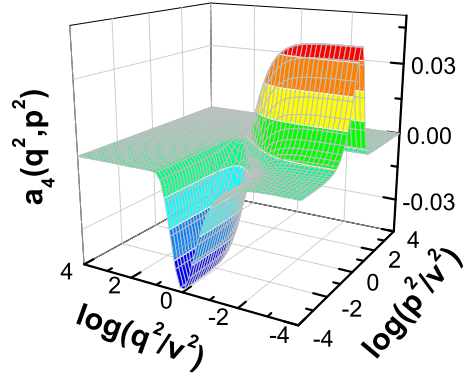}}
\vspace*{-3mm}
\caption{Calculated momentum dependence of the coefficient $a_{4}^{}$ (with the scale parameter being taken as $\textrm{v}=1.0\,$GeV).}
\label{fig:l4}
\end{figure}

The obtained momentum dependence of the coefficient $a_{4}^{}$ is shown in Fig.~\ref{fig:l4}.
It is evident that, even though the $a_{4}^{}$ oscillates as the momentums vary, the absolute value of
the $a_{4}^{}$ is quite small (lower than 0.05).
This fact accords with the constraint both from the WTI (as shown in Eq.(\ref{WTI})) and the STI in Eq.(\ref{STI}).

The momentum dependence of the transverse structure of the quark-gluon interaction vertex
has not yet been discussed in detail in the past decades,
even though some works have shown the importance of the transversal  part~\cite{DSE-BSE200912,CLR,QCLR,BCPR,Williams:2015EPJA,Williams:20156,Eichmann:2016PPNP,Eichmann:2018PRD,AW:2018CPC}.
In the following, we analyze the momentum dependence of the coefficients of the 8 transverse
Lorentz structures.

\begin{figure}[hbt]
\begin{minipage}[t]{0.5\textwidth}
\begin{minipage}{0.49\textwidth}
\centerline{\includegraphics[clip,width=\textwidth]{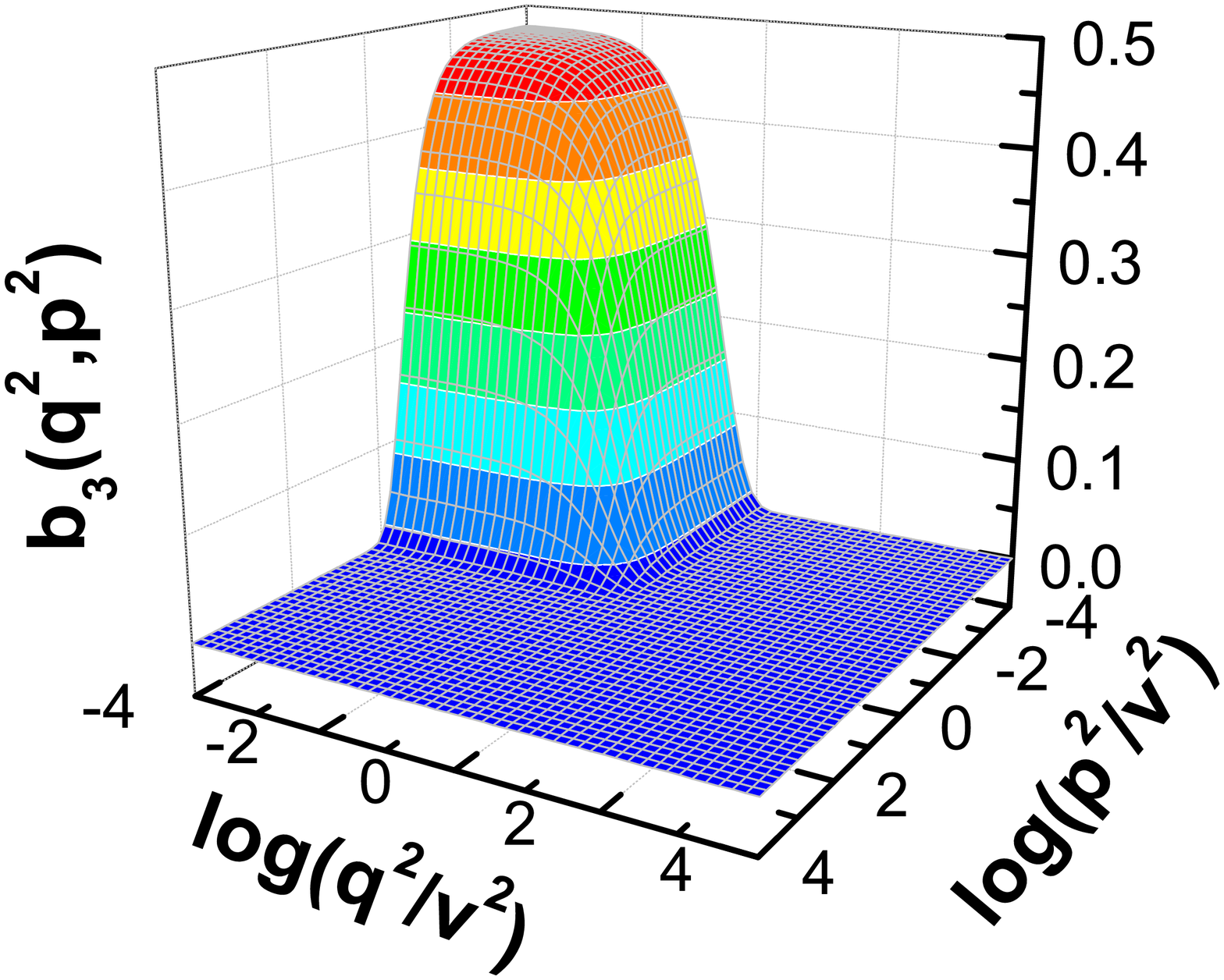}}
\end{minipage}
\begin{minipage}{0.49\textwidth}
\hspace*{-3ex}\includegraphics[width=\textwidth]{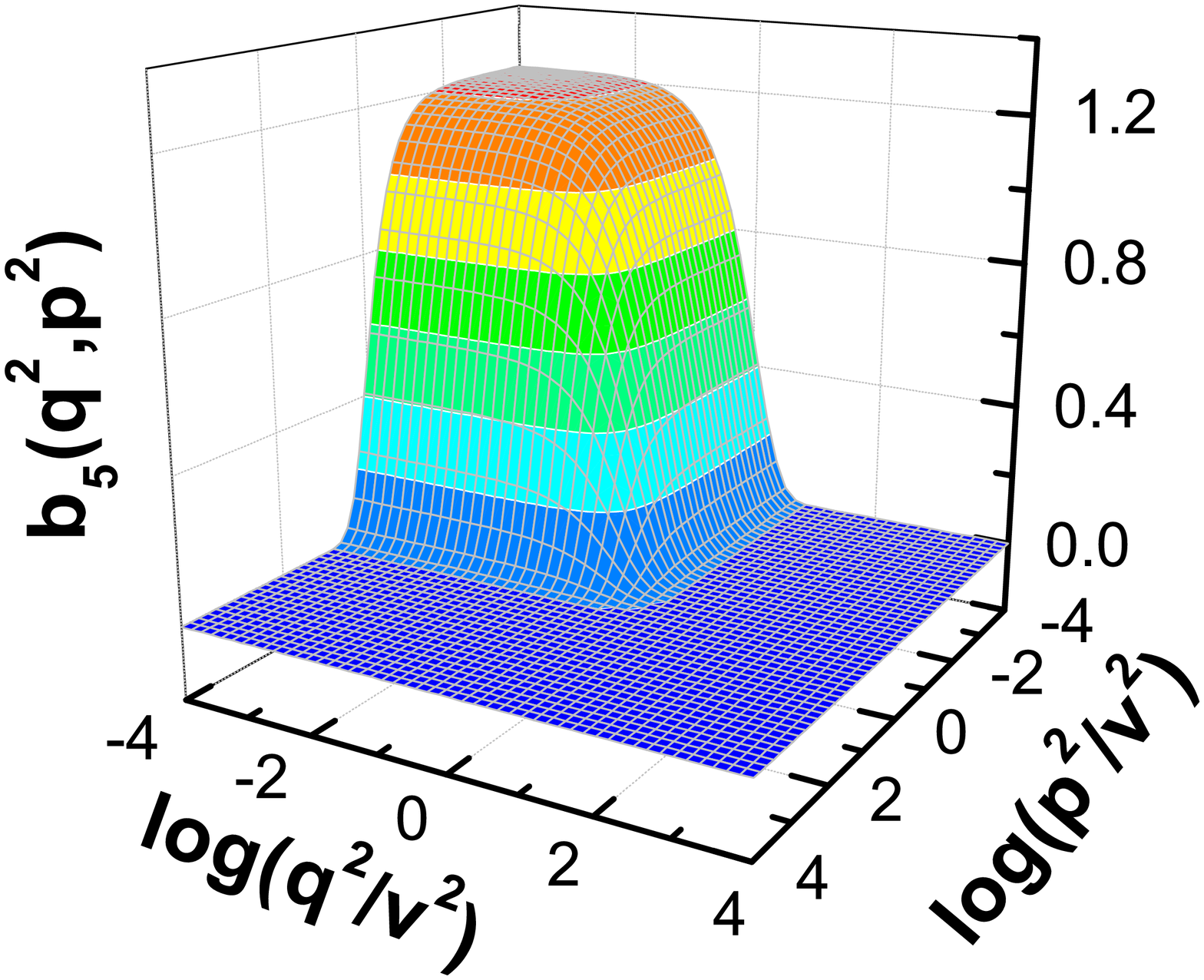}
\end{minipage}
\end{minipage}
\begin{minipage}[t]{0.5\textwidth}
\begin{minipage}{0.49\textwidth}
\centerline{\includegraphics[clip,width=\textwidth]{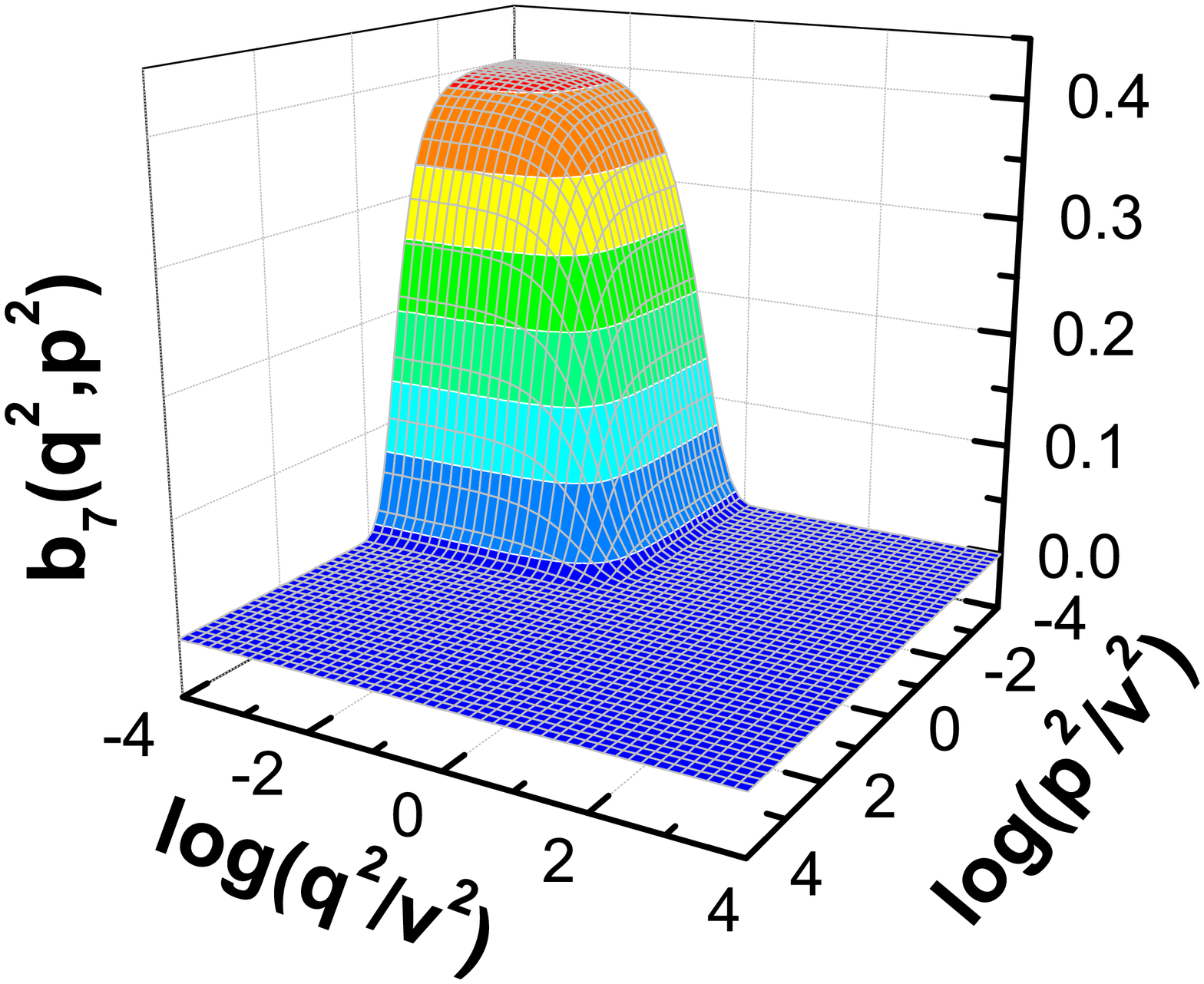}}
\end{minipage}
\begin{minipage}{0.49\textwidth}
\hspace*{-3ex}\includegraphics[width=\textwidth]{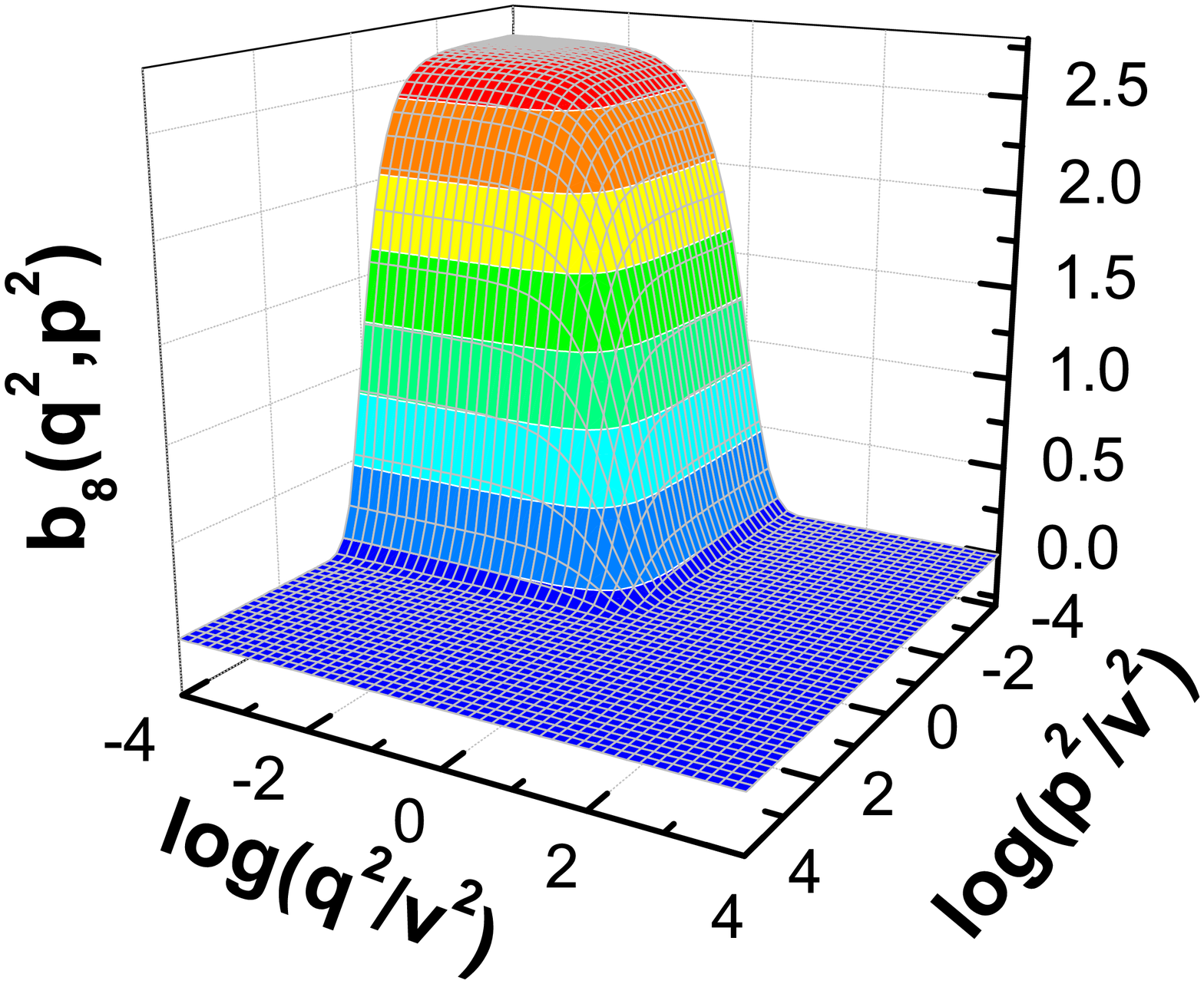}
\end{minipage}
\end{minipage}
\caption{Calculated momentum dependence of the coefficients $b_{3}^{}$ (\emph{upper-left panel}),
 $b_{5}^{}$ (\emph{upper-right panel}),  $b_{7}^{}$ (\emph{lower-left panel}) and $b_{8}^{}$
(\emph{lower-right panel}). The scale parameter is also set at $\textrm{v} = 1.0\,$GeV.}
\label{fig:t3t5t7t8}
\end{figure}

The calculated results of the momentum dependence of the coefficients $b_{3}^{}$, $b_{5}^{}$, $b_{7}^{}$ and $b_{8}^{}$ are displayed in Fig.\ref{fig:t3t5t7t8}.   %
%
%
From the Fig.~\ref{fig:t3t5t7t8}, one can notice easily that, at any momentum $q^{2}$ and $p^{2}$, the $b_{3}^{}$, $b_{5}^{}$, $b_{7}^{}$ and $b_{8}^{}$ take positive definite values.
And their  variation behaviors with respect to the momentums are quite similar: there is a flat-form in the IF region,
and the amplitudes reduce to approach 0 in the UV area.
However, the value of the amplitude is quantitatively different, specifically,
$b_{3}^{}(0,0)=0.48$, $b_{5}^{}(0,0)=1.22$, $b_{7}^{}(0,0)=0.4$, $b_{8}^{}(0,0)=2.63$.
This result matches qualitatively with the CLRQ vertex~\cite{CLR}:
 \begin{eqnarray}
\begin{split}
b_{5}^{}(q^{2},p^{2}) \! & \! = \! & \! \eta\Delta_{B}^{}(q^{2}, p^{2}) \, , \\
b_{8}^{}(q^{2},p^{2}) \! & \! = \! & \! \frac{2 a_{5}^{}(q^{2}, p^{2})}{\mathcal{M}(q^{2}, p^{2})} \, ,
\end{split}
\end{eqnarray}
where $\eta=-\frac{7}{4}$ and $\mathcal{M}(x,y)=[x+y+M(x)^2+M(y)^2]/(2[M(x)+M(y)])$,
and leads then to the similar behavior of the quark's anomalous magnetic moment as shown in next section.

\begin{figure}[hbt]
\begin{minipage}[t]{0.5\textwidth}
\begin{minipage}{0.49\textwidth}
\centerline{\includegraphics[clip,width=\textwidth]{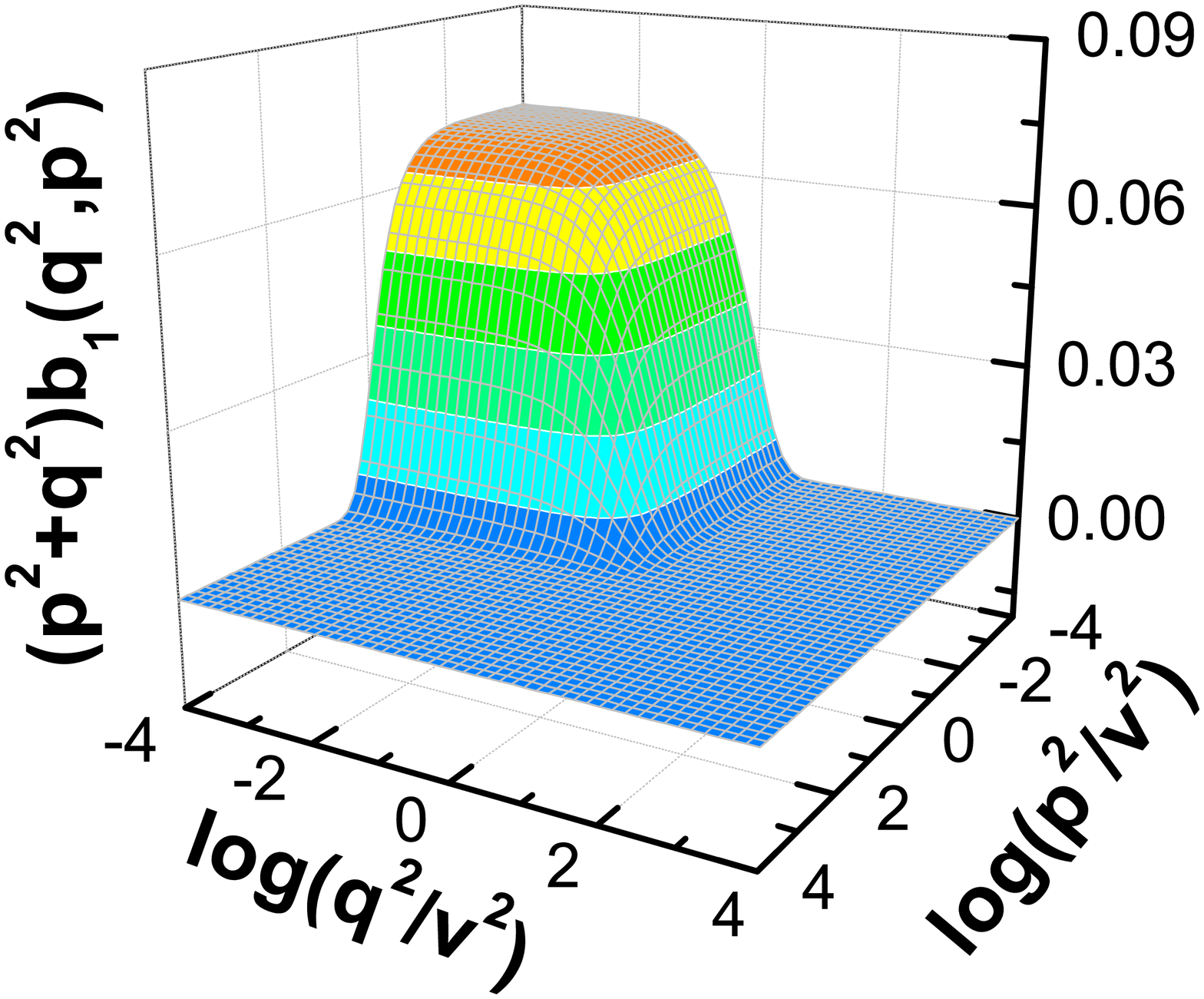}}
\end{minipage}
\begin{minipage}{0.49\textwidth}
\hspace*{-3ex}\includegraphics[width=\textwidth]{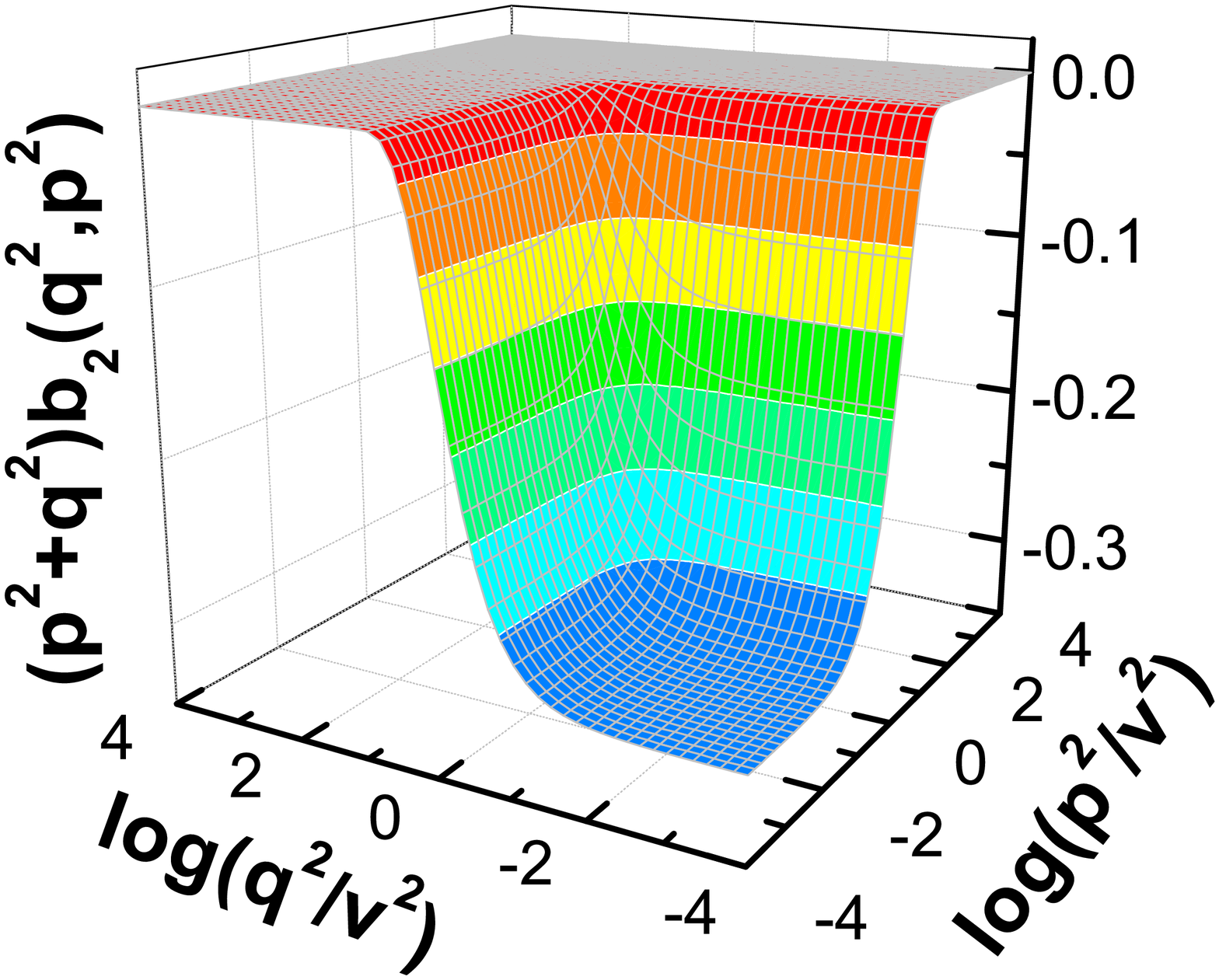}
\end{minipage}
\end{minipage}
\begin{minipage}[t]{0.5\textwidth}
\begin{minipage}{0.49\textwidth}
\centerline{\includegraphics[clip,width=\textwidth]{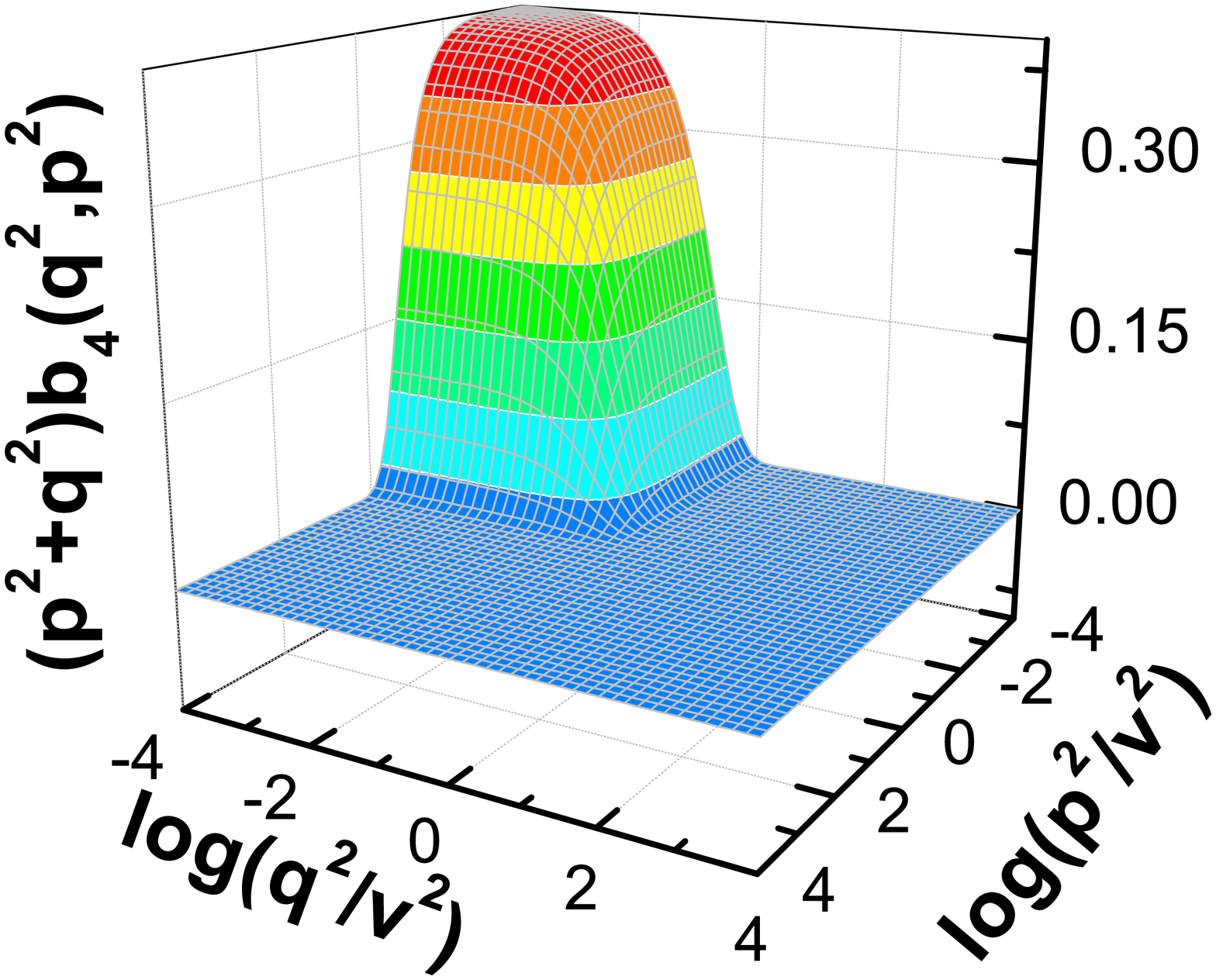}}
\end{minipage}
\begin{minipage}{0.49\textwidth}
\hspace*{-3ex}\includegraphics[width=\textwidth]{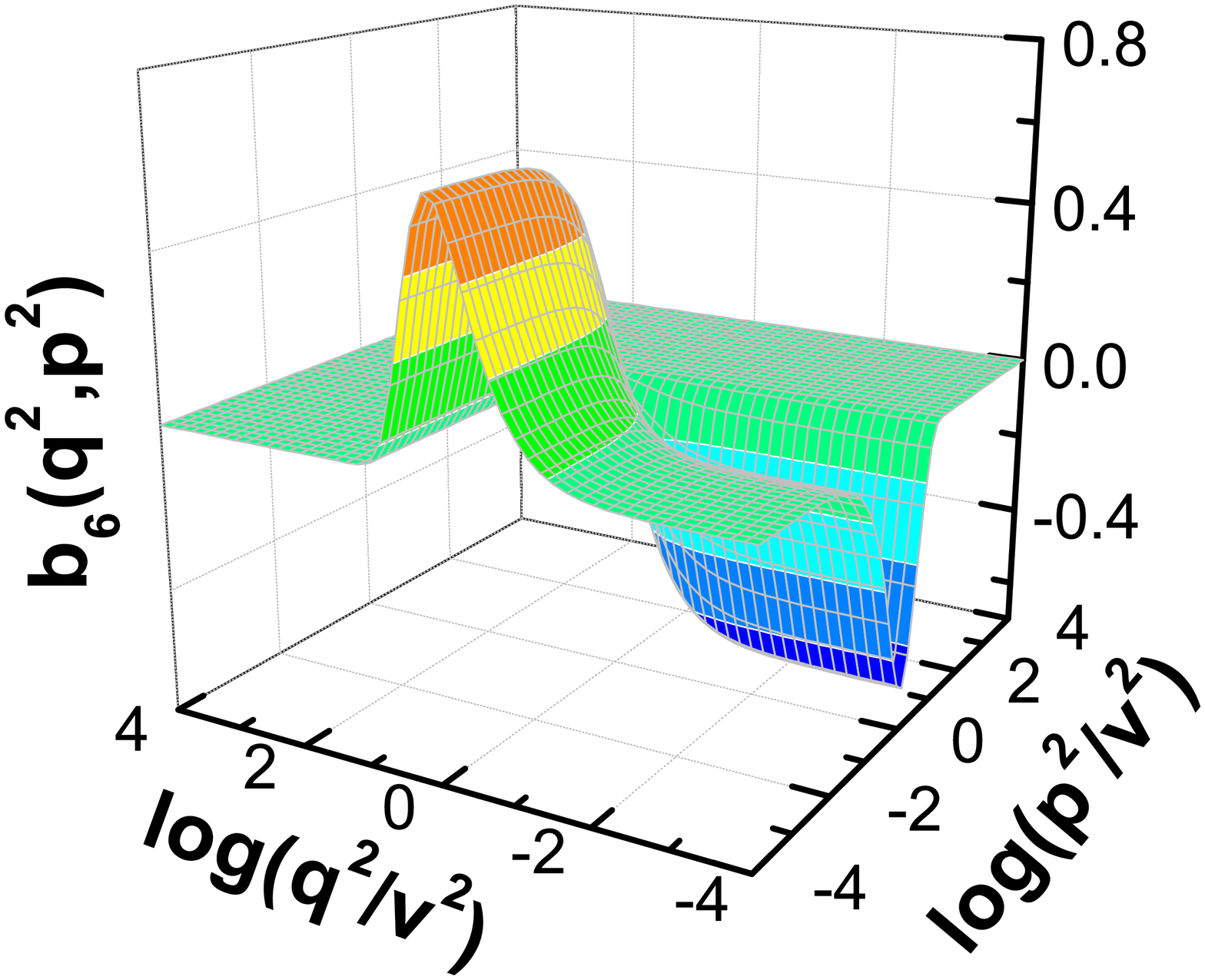}
\end{minipage}
\end{minipage}
\caption{Calculated momentum dependence of the coefficient $(p^{2} + q^{2})b_{1}^{}$ (\emph{upper-left panel}) ,
the coefficient $(p^{2} + q^{2})b_{2}^{}$ (\emph{upper-right panel}),
the coefficient $(p^{2} + q^{2})b_{4}^{}$ (\emph{lower-left panel}),
and the coefficient $b_{6}^{}$ (\emph{lower-right panel}).
The scale parameter is also taken as $\textrm{v}=1.0\,$GeV.}
\label{fig:t1t2t4t6}
\end{figure}

The obtained results of the momentum dependence of the coefficients $b_{1}^{}$, $b_{2}^{}$, $b_{4}^{}$ and $b_{6}^{}$ are illustrated in Fig.~\ref{fig:t1t2t4t6}.
For the practical convenience of removing the divergence at $p^{2} + q^{2} =0$ in intuitive view in figure,
we show the  behaviors of the  $b_{1}^{}$, $b_{2}^{}$ and $b_{4}^{}$ with a factor $(p^{2} + q^{2})$ being multiplied in Fig.~\ref{fig:t1t2t4t6}.
 One can observe from Fig.~\ref{fig:t1t2t4t6} that,
 similar to the behaviors of $b_{3}^{}$, $b_{5}^{}$, $b_{7}^{}$ and $b_{8}^{}$, at any momentum $q^{2}$ and $p^{2}$, the $(p^{2} + q^{2}) b_{1}^{}$  and $(p^{2} + q^{2})b_{4}^{}$ hold positive definite values, but the $(p^{2} + q^{2}) b_{2}^{}$ has negative value.
They also have their own similar  behaviors with respect to the momentums: there is a flat-form in the IF region for each of the $(p^{2} + q^{2}) b_{1}^{}$, $(p^{2} + q^{2}) b_{2}^{}$ and $(p^{2} + q^{2}) b_{4}^{}$,
and their amplitudes (the absolute values) fall quickly to 0 as the momentums get larger.
 The infrared finite behaviors also reveal that these three coefficients tend to $1/(p^2+q^2)$
divergence when $p^{2} + q^{2}$ goes to zero.
These behaviors agree with that given in Refs. [39,40] (after transformation and considering the approximation we made on the angle $\overline{\theta} = 0.49 \pi$, one can find that the coefficients $h_{2}$ and $h_{3}$ approximate to the $a_{2} + 2(p^{2} + q^{2})b_{2}$ and $-2a_{3} + (p^{2} + q^{2})b_{1}$ in our work).
As for the coefficient $b_{6}$, its  behavior is qualitatively the same as that of the coefficient $a_{4}$,  except for that its magnitude is much larger.

Besides, in general, analyzing the tensor structure of our basis of the vertex, one can notice that the $L^{4}$ and the $T^{6}$ have odd C-parity, while others have even C-parity. Since the C-parity is conserved in QCD, the vertex dressing functions $a_{4}$ and $b_{6}$ should be odd in  $(p^2-q^2)$ and the others should be even. Our presently obtained results match these general features well.

\subsubsection{Current quark mass dependence of the vertex}

In this part, we discuss the current quark mass dependence of the 12 coefficients
with respect to the momentum.
For easy access, we here just show the single momentum dependence $a_{i}^{}(0,p^2)\;\; (i = 1,2,3,4)$ and $b_{j}^{}(0,p^2)\;\; (j = 1,2, \cdots, 8)$ in case of several values of the current quark mass.
While the 3-dimensional variation behavior of the quark-gluon vertex with non-vanishing current mass $m_{f}^{}$
delivers certainly the similar behavior as that in case of the chiral limit.

\begin{figure}[hbt]
\begin{center}
\includegraphics[width=0.46\textwidth]{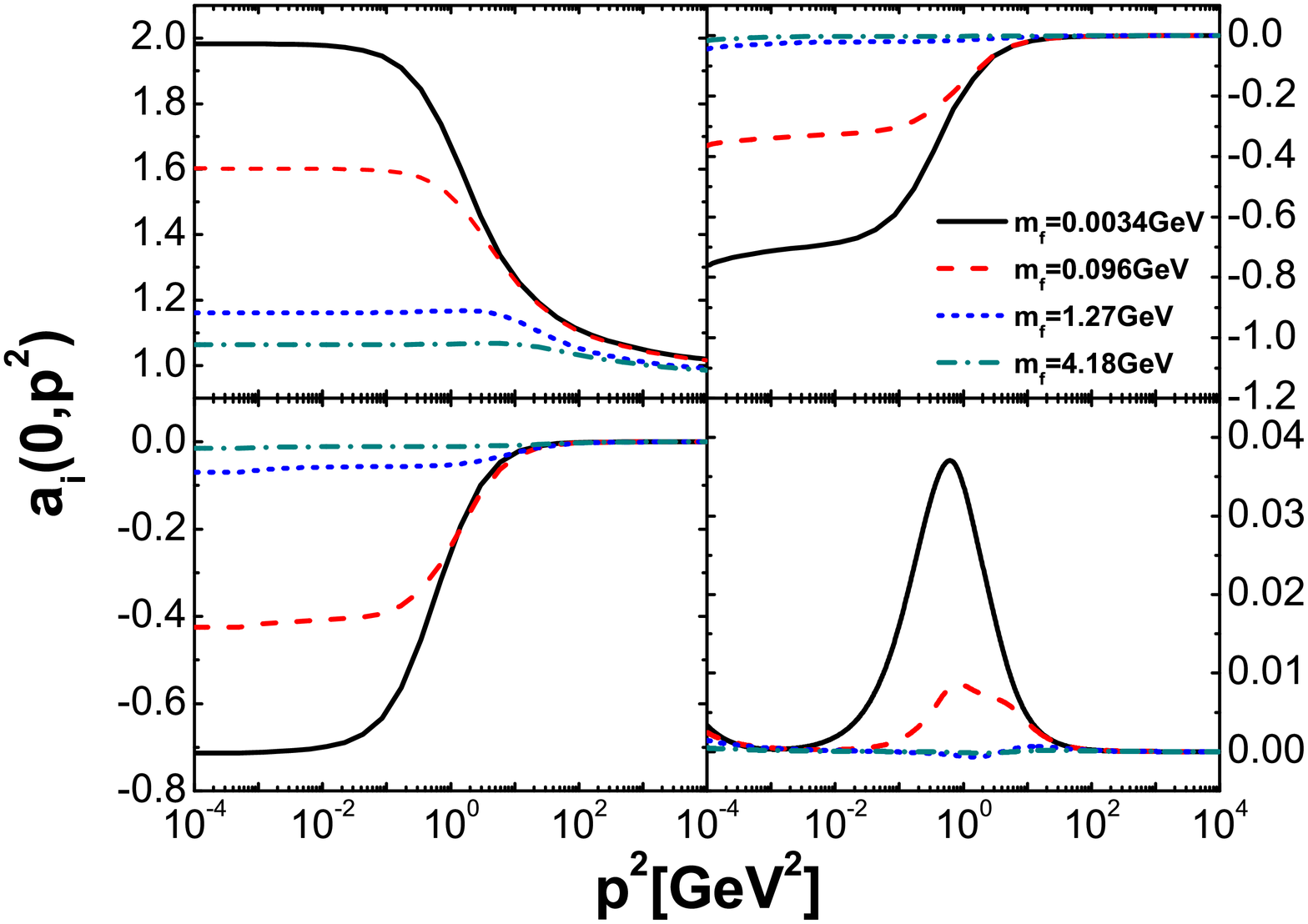}
\end{center}
\vspace*{-8mm}
\caption{Calculated momentum dependence of the longitudinal structures's coefficients at several values of the  current quark mass.
\emph{Upper-Left}: the coefficient $a_{1}^{}$,
\emph{Upper-Right}: the coefficient $a_{2}^{}$,
\emph{Lower-Left}: the coefficient $a_{3}^{}$,
\emph{Lower-Right}: the coefficiet $a_{4}^{}$.}
\label{fig:a1a2a3a4}
\end{figure}

Calculated momentum dependence of the longitudinal structures' coefficients is illustrated in Fig.~\ref{fig:a1a2a3a4}.
It is worth noticing that all the amplitudes (the absolute values) of the four coefficients are monotonously suppressed as the current quark mass increases.
As for the momentum dependence, the $a_{i}^{}(0, p^{2})\;\; (i = 1 ,2,3)$ vary with respect to $p^{2}$ monotonously, which  is different from that given in BC vertex (in which $a^{BC}_2=\Delta A(p^2)$,
while the $A(p^2)$ from the BC vertex is not monotonous),
but consistent with the results from the lattice QCD simulations~\cite{SBKOSSW}, qualitatively.
The reason might be that the WTI is not a correct constraint to the vertex any more,
after taking the non-Abelian nature of the interaction into consideration.
However,  it is exciting to see that the curves are in accordance with the STI results~\cite{ACFP} qualitatively.
Another remarkable feature is that, in all the cases of the current quark masses,
the contribution of the $L^{4}$ to the full vertex is quite small.

\begin{figure}[hbt]
\begin{center}
\includegraphics[width=0.45\textwidth]{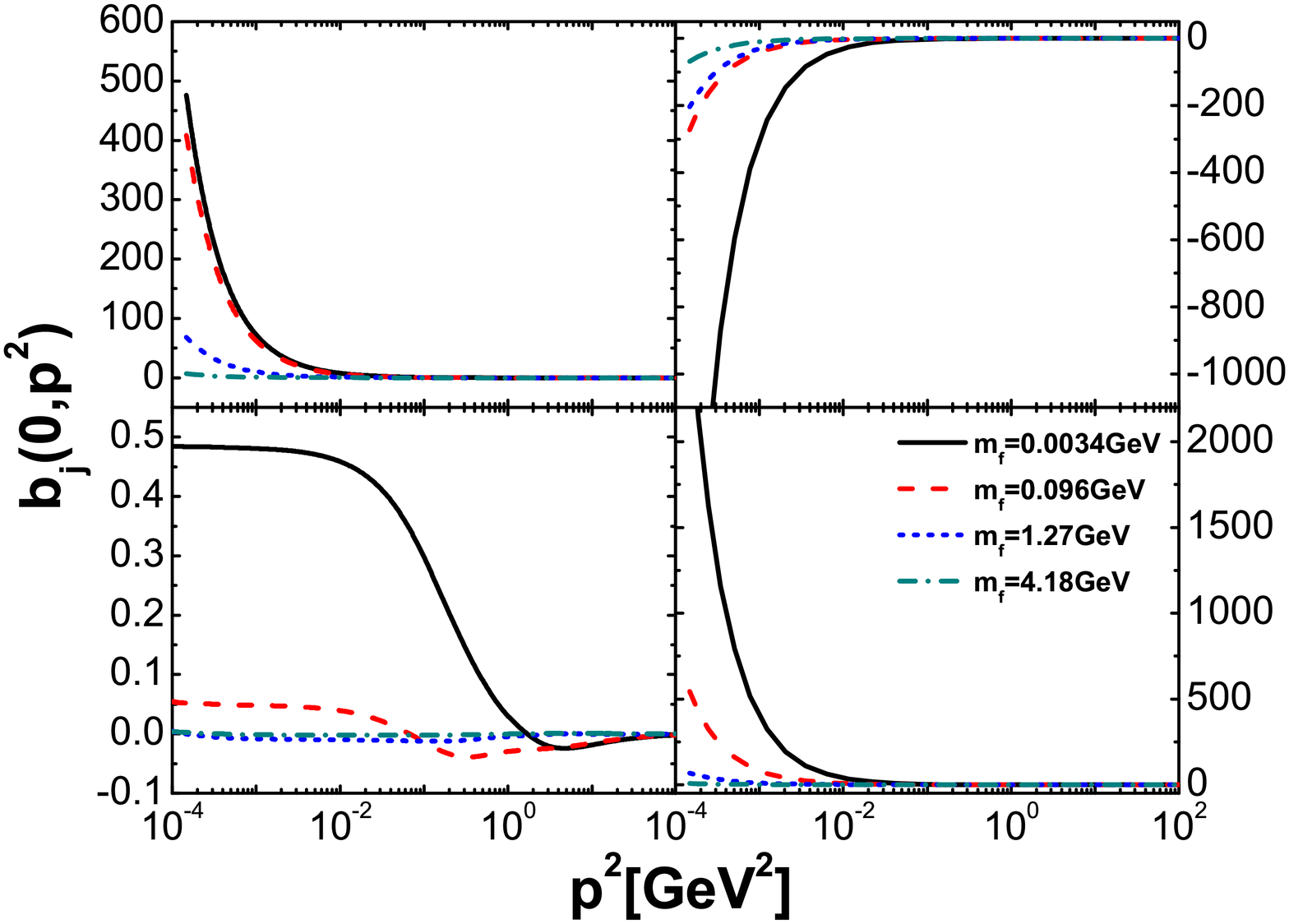}
\end{center}
\vspace*{-6mm}
\caption{Calculated momentum dependence of the transverse structures's coefficients $b_{1}^{}(0, p^{2})$,
 $b_{2}^{}(0, p^{2})$,  $b_{3}^{}(0, p^{2})$ and $b_{4}^{}(0, p^{2})$ at several values of the current mass.
\emph{Upper-Left}: the coefficient $b_{1}^{}$,
\emph{Upper-Right}: the coefficient $b_{2}^{}$,
\emph{Lower-Left}: the coefficient $b_{3}^{}$,
\emph{Lower-Right}: the coefficient $b_{4}^{}$. }
\label{fig:b3b5b7b8}
\end{figure}

\begin{figure}[hbt]
\begin{minipage}{0.46\textwidth}
\centerline{\includegraphics[clip,width=\textwidth]{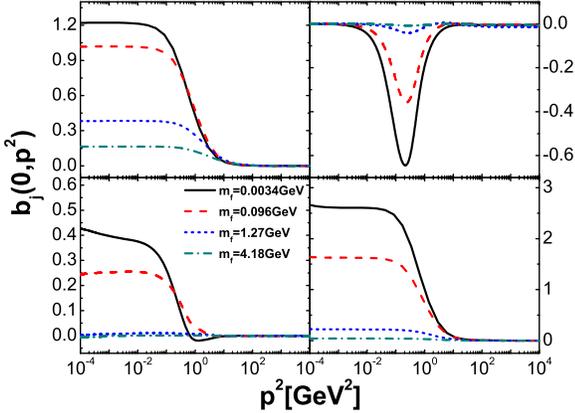}}
\end{minipage}
\vspace*{-3mm}
\caption{Calculated momentum dependence of the transverse structures's coefficients $b_{5}^{}(0, p^{2})$,
 $b_{6}^{}(0, p^{2})$,  $b_{7}^{}(0, p^{2})$ and $b_{8}^{}(0, p^{2})$ at several values of the current mass.
\emph{Upper-Left}: the coefficient $b_{5}^{}$,
\emph{Upper-Right}: the coefficient $b_{6}^{}$,
\emph{Lower-Left}: the coefficient $b_{7}^{}$,
\emph{Lower-Right}: the coefficient $b_{8}^{}$ . }
\label{fig:b1b2b4b6}
\end{figure}

The calculated momentum dependence of the transverse structures' coefficients in case of several values of
the current quark mass are shown in Fig.~\ref{fig:b3b5b7b8} and Fig.~\ref{fig:b1b2b4b6}.
The figures clearly manifest  that the $b_{1}^{}$, $b_{2}^{}$ and $b_{4}^{}$ fall quite rapidly to 0
with the increasing of both the momentum and the current mass.
The $b_{3}^{}$, $b_{5}^{}$, $b_{7}^{}$ and $b_{8}^{}$, especially the $b_{5}^{}$ and $b_{8}^{}$ for the $u$-, $d$- and
$s$-quarks, hold considerable values in the IF region. While the value of $b_{6}^{}$ keeps small,
and the maximum of the peak appearing at mediate-momentum region is only $0.65$.
It indicates that the magnitudes of the coefficients of the $T^{1}$, $T^{2}$, $T^{4}$ and $T^{6}$ terms
are small and  can usually be neglected.
However the coefficients of the $T_{5}^{}$ and $T_{8}^{}$ terms
which are included in the CLRQ vertex~\cite{CLR,QCLR}, are large enough in the infrared area,
especially for the $u$-, $d$- and $s$-quark.
This provides another positive evidence for the CLRQ model of the quark-gluon interaction vertex.
Meanwhile, the values of the coefficients of the $T_{3}^{}$ and $T_{7}^{}$ terms can not be ignored for light quark, either.

It is worth noticing  that in  Figs.~\ref{fig:b3b5b7b8} and \ref{fig:b1b2b4b6},
generally, with the increasing of the current quark mass, the contributions of the transverse parts to the full vertex
get weakened drastically. Eventually, the transverse parts can be ignored for very heavy quark (e.g., $b$-quark) system~\cite{DGCL}.

\section{Parameterizing the Quark-gluon interaction vertex}   \label{Sec:BuildingNewScheme}

To make use of the presently determined quark-gluon interaction vertex elsewhere,
it is necessary to parameterize the above results as analytical functions.
We then model the vertex with functions including some parameters.
After some tedious calculations, the coefficients of the 12 Lorentz structures can be fitted as:
\begin{eqnarray}\label{model}
\begin{split}
a_{i}^{}(p,q) = c^{a}_{i}(p^{2},q^{2})e^{-(p^{2}+q^{2})/d^{a}_{i}(m)}   \, , \\
b_{j}^{}(p,q) = c^{b}_{j}(p^{2},q^{2})e^{-(p^{2}+q^{2})/d^{b}_{j}(m)} \, ,
\end{split}
\end{eqnarray}
with $i = 1,2,3,4$, and $j = 1,2, \cdots, 8$.

The current quark mass effect can also be modeled in analytic expression of the vertex.
The fitted expressions can be written explicitly as:
\begin{eqnarray}\label{model11}
\begin{split}
a_{1}(q^2,p^2) = & \; 1.0+\Big{(} 0.058+\frac{0.136}{m_f+0.15} \Big{)}   \\
 & \displaystyle \times e^{-(q^2+p^2)/(379-\frac{2446}{m_f+6.49})} \, , \\
a_{2}(q^2,p^2) = & \, \Big{(} 0.027-\frac{0.06}{m_f+0.08} \Big{)}         \\
 & \displaystyle  \times e^{-(q^2+p^2)/(5.84-\frac{1.68}{m_f+0.314})}  \, , \\
a_{3}(q^2,p^2) = & \, \Big{(} 0.013 - \frac{0.095}{m_f+0.133} \Big{)}  \\
 & \displaystyle \times e^{-(q^2+p^2)/(55.5-\frac{264}{m_f+4.83})} \, ,  \\
a_{4}(q^2,p^2) = & \, \Big{(} 0.0056 - \frac{0.0034}{m_f+0.026} \Big{)} \\
 & \displaystyle \times (q^2 - p^2)e^{-(p^2+q^2)} \, , \\
b_{1}(q^2,p^2) = & \, \frac{0.026+\frac{0.0007}{m_f+0.016}}{q^2+p^2} \\
 & \displaystyle \times e^{-(q^2+p^2)/(55.5-\frac{264}{m_f+4.83})} \, ,  \\
b_{2}(q^2,p^2) = & \, -\frac{0.014+\frac{0.0016}{m_f+0.006}}{q^2+p^2}  \\
 & \displaystyle  \times e^{-(q^2+p^2)/(5.84-\frac{1.68}{m_f+0.314})}  \, , \\
\end{split}
\end{eqnarray}
\begin{eqnarray}\label{model12}
\begin{split}
b_{3}(q^2,p^2) = & \, \Big{(} -0.029+\frac{0.011}{m_f+0.021} \Big{)}  \\
 & \displaystyle \times e^{-(q^2+p^2)/(-0.015+\frac{0.007}{m_f+0.026})} \, , \\
b_{4}(q^2,p^2) = & \frac{0.078+\frac{0.0008}{m_f+0.002}}{q^2+p^2} \\
 & \displaystyle \times e^{-(q^2+p^2)/(1.53-\frac{0.53}{m_f+0.41})} \, , \\
b_{5}(q^2,p^2) = & \Big{(} 0.062 +\frac{0.52}{m_f+0.44} \Big{)} \\
 & \displaystyle \times e^{-(q^2+p^2)/(4.52-\frac{2.42}{m_f+0.7})} \, , \\
b_{6}(q^2,p^2) = & \Big{(} - 0.073 + \frac{0.127}{m_f+0.077} \Big{)}  \\
 & \displaystyle \times (q^2 - p^2)e^{-(q^2+p^2)} \, , \\
b_{7}(q^2,p^2) = & \big{(} -0.036 + \frac{0.067}{m_f+0.14} \Big{)}  \\
 & \displaystyle \times e^{-(q^2+p^2)/(1.53-\frac{0.53}{m_f+0.41})} \, , \\
b_8(q^2,p^2) = & \Big{(} -0.075 + \frac{0.43}{m_f+0.16} \Big{)}  \\
 & \displaystyle \times e^{-(q^2+p^2)/(3.72-\frac{2.3}{m_f+0.88})} \, .
\end{split}
\end{eqnarray}
With the expression  of the gluon and the running coupling described in
Section~\uppercase\expandafter{\romannumeral3},
a practical truncation scheme of solving the DSE of the quark propagator can be built.

\begin{figure}[hbt]
\begin{center}
\includegraphics[width=0.41\textwidth]{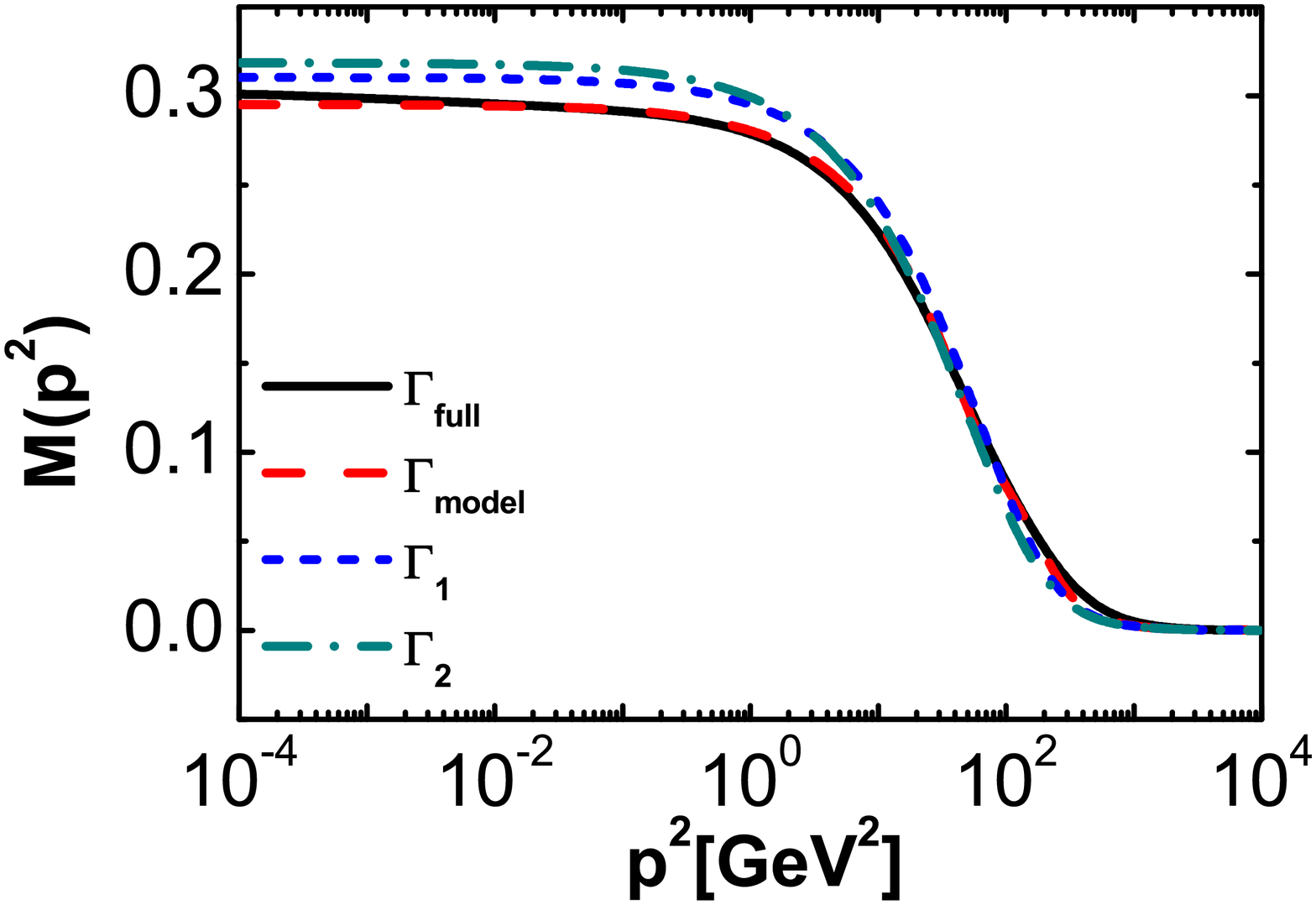}
\vspace*{-5mm}
\includegraphics[width=0.41\textwidth]{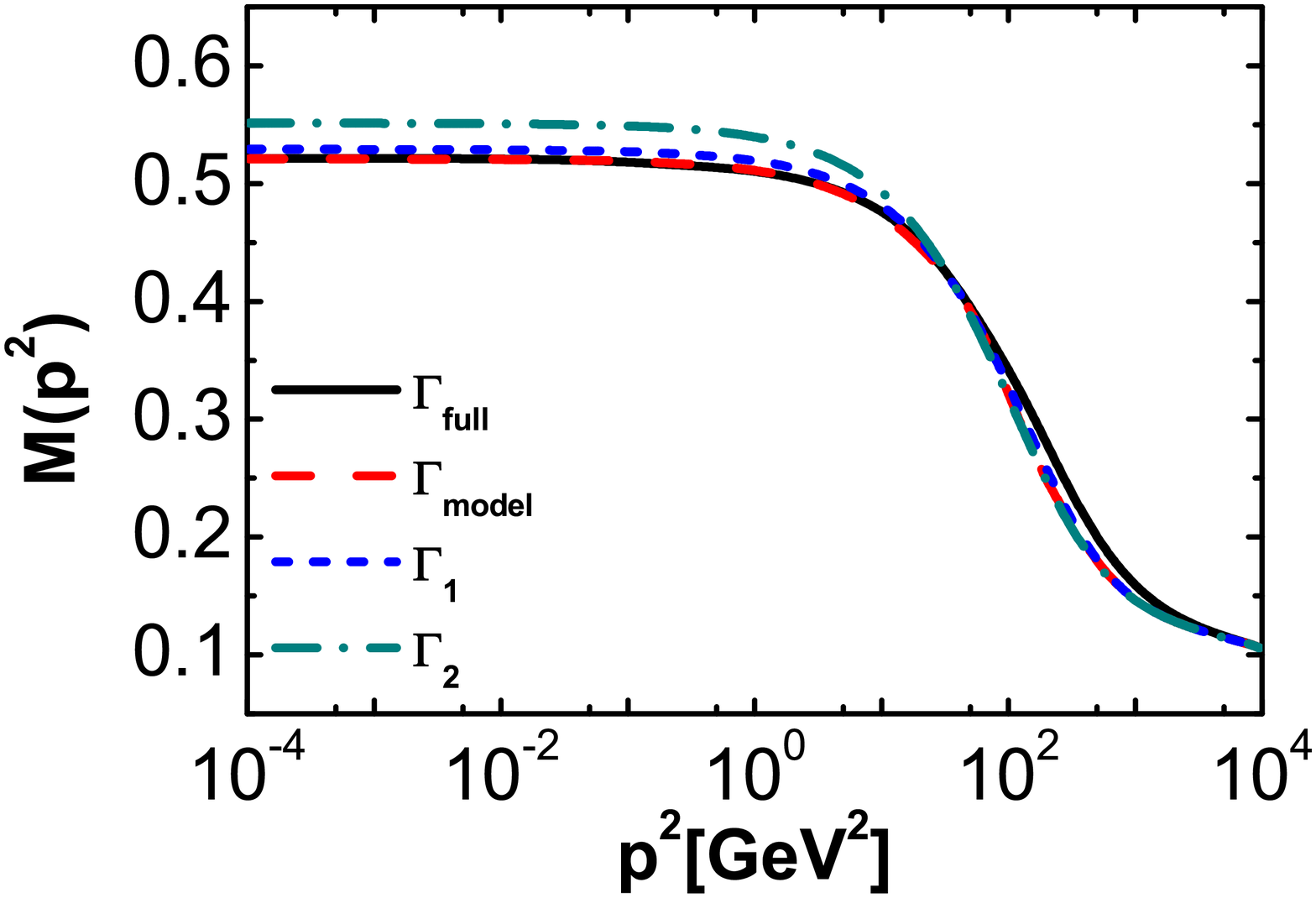}
\end{center}
\vspace*{-2mm}
\caption{Calculated dynamical mass (in unit GeV) of the quark in and beyond the chiral limit via different vertexes.
The one marked as $\Gamma_{\textrm{full}}$ is the result obtained via the vertex directly from the coupled DSEs,
The $\Gamma_{\textrm{model}}$ is the result via the above parameterized expression of the vertex,
The $\Gamma_{1}$ is that through the simplifies vertex which includes the Lorentz structures
$L^{1}$, $L^{2}$, $L^{3}$, $T^{5}$ and $T^{8}$,
The $\Gamma_{2}$ is that coming from the further simplified model composed of only
the Lorentz structure $L^{1}$, $T^{5}$ and $T^{8}$.
\emph{Upper Panel}: in case of the chiral limit;
\emph{Lower Panel}: in case beyond the chiral limit with $m_{f}^{}=96\,$MeV. }
\label{fig:diffvertex}
\end{figure}

To check the efficiency of the presently built truncation scheme of the DSEs,
we calculated the dynamical mass of the quark in and beyond the chiral limit with the
full parameterized expression of the vertex listed above,
the parameterized vertex with simplification so as to include the Lorentz structures
$L^{1}$, $L^{2}$, $L^{3}$, $T^{5}$ and $T^{8}$,
and the further simplified model including only the $L^{1}$, $T^{5}$ and $T^{8}$ terms.
The obtained results and the comparison with that coming from the full vertex directly from the coupled DSEs
are displayed in Fig.\ref{fig:diffvertex}.
The figure manifests clearly that the dynamical mass obtained via the parameterized expression of
the full vertex is very close to that straightly from the vertex given by the coupled DSEs.
We also find that the contributions from the Lorentz structure $L^{1}$, $L^{2}$, $L^{3}$, $T^{5}$
and $T^{8}$ (just the same as in the CLRQ vertex~\cite{CLR}) are dominant for obtaining the same DCSB
effect in quark propagator.

\section{Application examples of the practical truncation method} \label{Sec:Applications}

\subsection{Quark-photon vertex}

\begin{figure}[hbt]
\centerline{\includegraphics[width=0.46\textwidth]{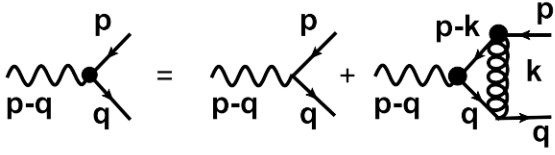}}
\vspace*{-4mm}
\caption{The BSE of the quark-photon vertex, the bolded vertex and all the propagators are the fully dressed ones.}
\label{fig:BSE}
\end{figure}

The quark-photon vertex can be obtained by solving the Bethe-Salpeter Equation (BSE)
with the vector channel interaction. Once the quark-gluon interaction vertex and the quark-quark scattering kernel are available, the BSEs can be solved. As described in Refs.~\cite{Mun,CJT}, the BS kernel can be obtained through  the variation of the self energy:
\begin{equation}
\label{eq:kernel}
K(p,q;0)=-\frac{\delta\Sigma(p)}{\delta S(q)} \, .
\end{equation}
 Since we now have the parameterized quark-gluon vertex, the modeling running coupling and the modeling gluon propagator,  and in this scheme we  approximately neglect their dependence on the functional of the quark propagator, which induces
\begin{equation}\label{app}
  \frac{\delta\Gamma_\nu(q,p)}{\delta S(q)}=\frac{\delta \alpha((p-q)^2)}{\delta S(q)}=\frac{\delta G_{\mu\nu}(p-q)}{\delta S(q)}=0\,,
\end{equation}
 the BSE of the quark-photon system can then be simplified as shown in Fig.~\ref{fig:BSE}.

In general, such a quark-photon interaction vertex can still be written as
\begin{equation} \label{tensors1-photon}
\Gamma^{\gamma}_\mu(q,p)=\Gamma^{\gamma,L}_\mu(q,p)+\Gamma^{\gamma,T}_\mu(q,p) \, ,
\end{equation}

where
\begin{equation}\label{tensors1L-photon-general}
\Gamma^{\gamma,L}_{\mu}(q,p) = \sum_{i=1}^{4} c_{i}^{}(q^{2} , p^{2}) L_{\mu}^{i}(p+q , p-q) \, ,
\end{equation}
\begin{equation}  \label{rensor1T-photon}
\Gamma^{\gamma,T}_{\mu}(q,p) = \sum^8_{i=1}d_{i}^{} (q^2,p^2)T^{i}_{\mu}(p+q,p-q) \, .
\end{equation}
We compute the solution of the gap equation with the parameterized form of the presently obtained quark-gluon interaction vertex, the gluon propagator and the running coupling described in Section~\uppercase\expandafter{\romannumeral3} being substituted into, and the same ansatz for the BSE as implemented in the gap equation.
We obtain, in turn, the dressing functions $c_{i}^{}$ $(i=1,2,3,4)$ and $d_{i}^{}\;\; (i=1,2,\cdots, 8)$.

One knows, in general, that the truncation of the BS equation and the DS equations breaks the symmetry of the system, and the longitudinal part should satisfy the WTI if the symmetry is preserved.
With the constraint from the WTI  being taken into account, the longitudinal part of the quark-photon vertex
takes the similar form as the Ball-Chiu ansatz, i.e., the $c_{i}^{}$ in the above general expression can be written directly as the superposition of the dressing functions in the quark propagator, which reads
\begin{eqnarray}\label{tensors1L-photon}
\begin{split}
c_{1}^{} =  & \; \Sigma_{A}^{}(q^{2} , p^{2}) = \frac{A(q^2) + A(p^2)}{2} \, ,  \\[1mm]
c_{2}^{} =  & \; \Delta_{A}^{} (q^{2}, p^{2}) = \frac{A(q^2) - A(p^2)}{q^2 - p^2}  \, ,   \\[1mm]
c_{3}^{} =  & \; \Delta_{B}^{} (q^{2}, p^{2}) = \frac{B(q^2) - B(p^2)}{q^2 - p^2} \, , \\[1mm]
c_{4}^{} \equiv  & \; 0  \, . \\
\end{split}
\end{eqnarray}
%
%

\begin{figure}[hbt]
\begin{center}
\includegraphics[width=0.42\textwidth]{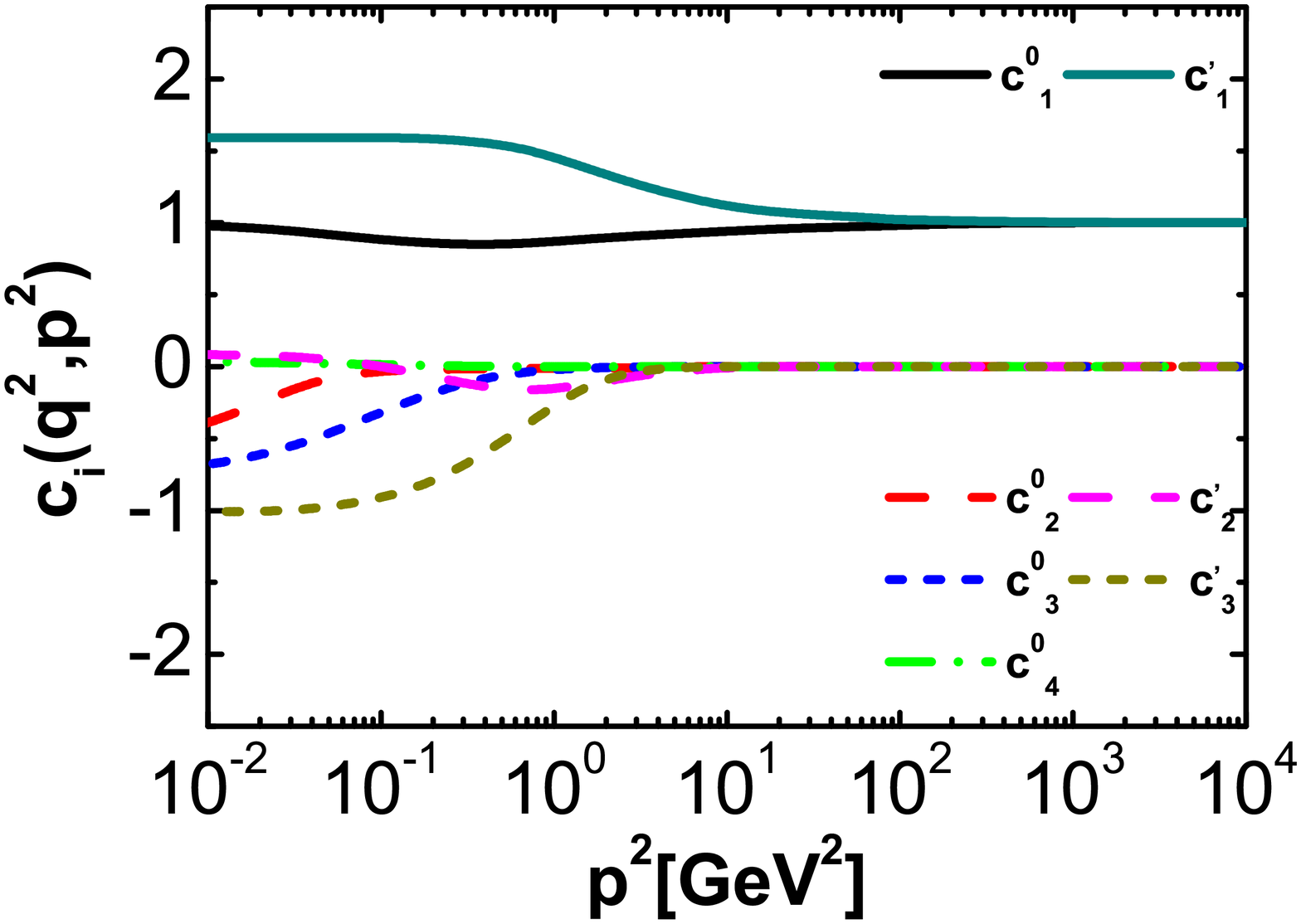}
\vspace*{-6mm}
\includegraphics[width=0.42\textwidth]{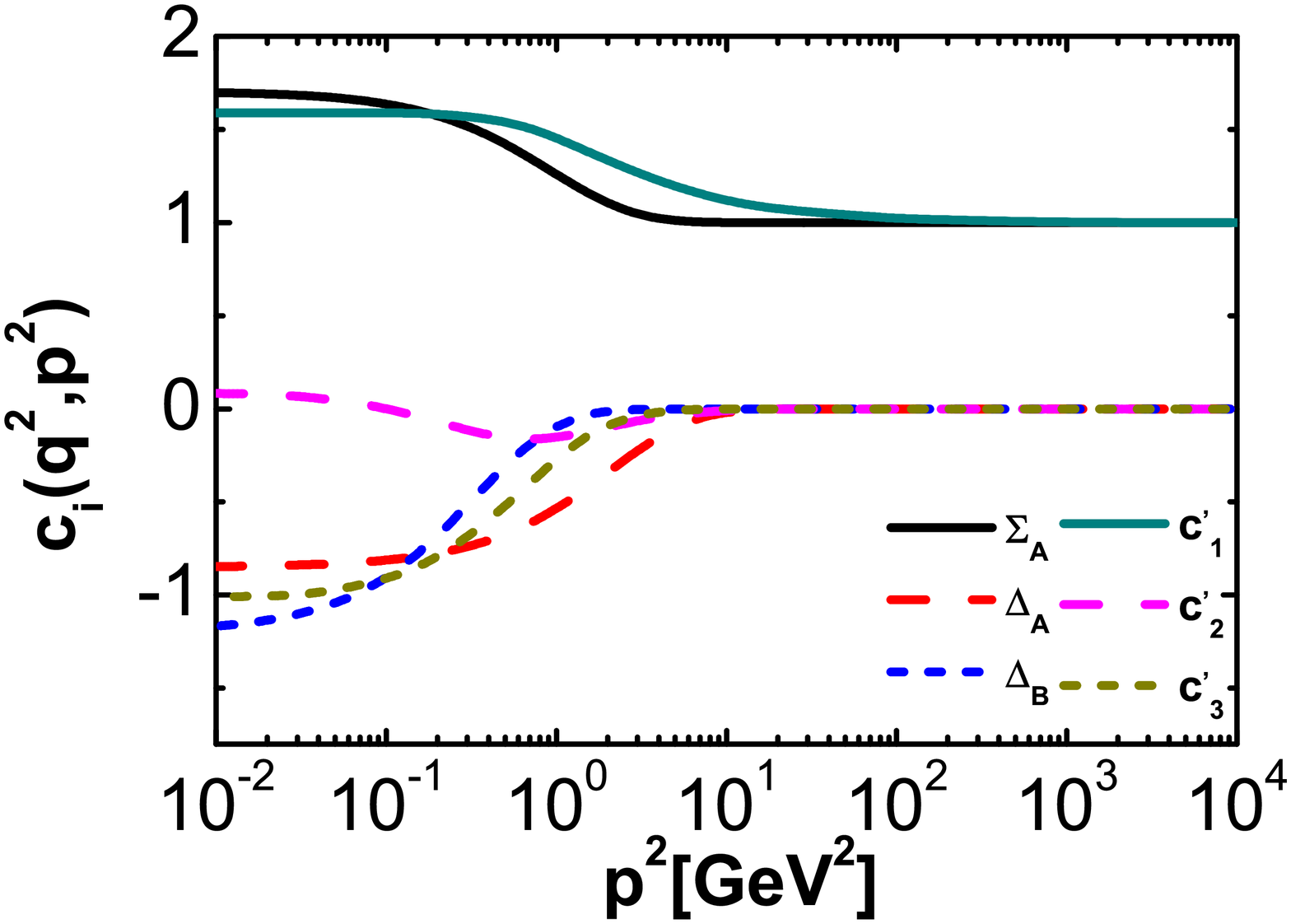}
\end{center}
\vspace*{-2mm}
\caption{Calculated momentum dependence of the longitudinal structures's coefficients
of the quark-photon vertex. \emph{Upper Panel}: without the constraint from the WTI;
\emph{Lower Panel}: with the constraint from the WTI.
The ones marked with $\Sigma_{A}^{}$, $\Delta_{A}^{}$ and $\Delta_{B}^{}$
are the results from the presently obtained quark-gluon interaction vertex and the constraint of the WTI,
while the $c_{1}^{\prime}$, $c_{2}^{\prime}$ and $c_{3}^{\prime}$ are those from the RL approximation.
 }
\label{fig:qqp1}
\end{figure}

\begin{figure}[hbt]
\begin{center}
\includegraphics[width=0.42\textwidth]{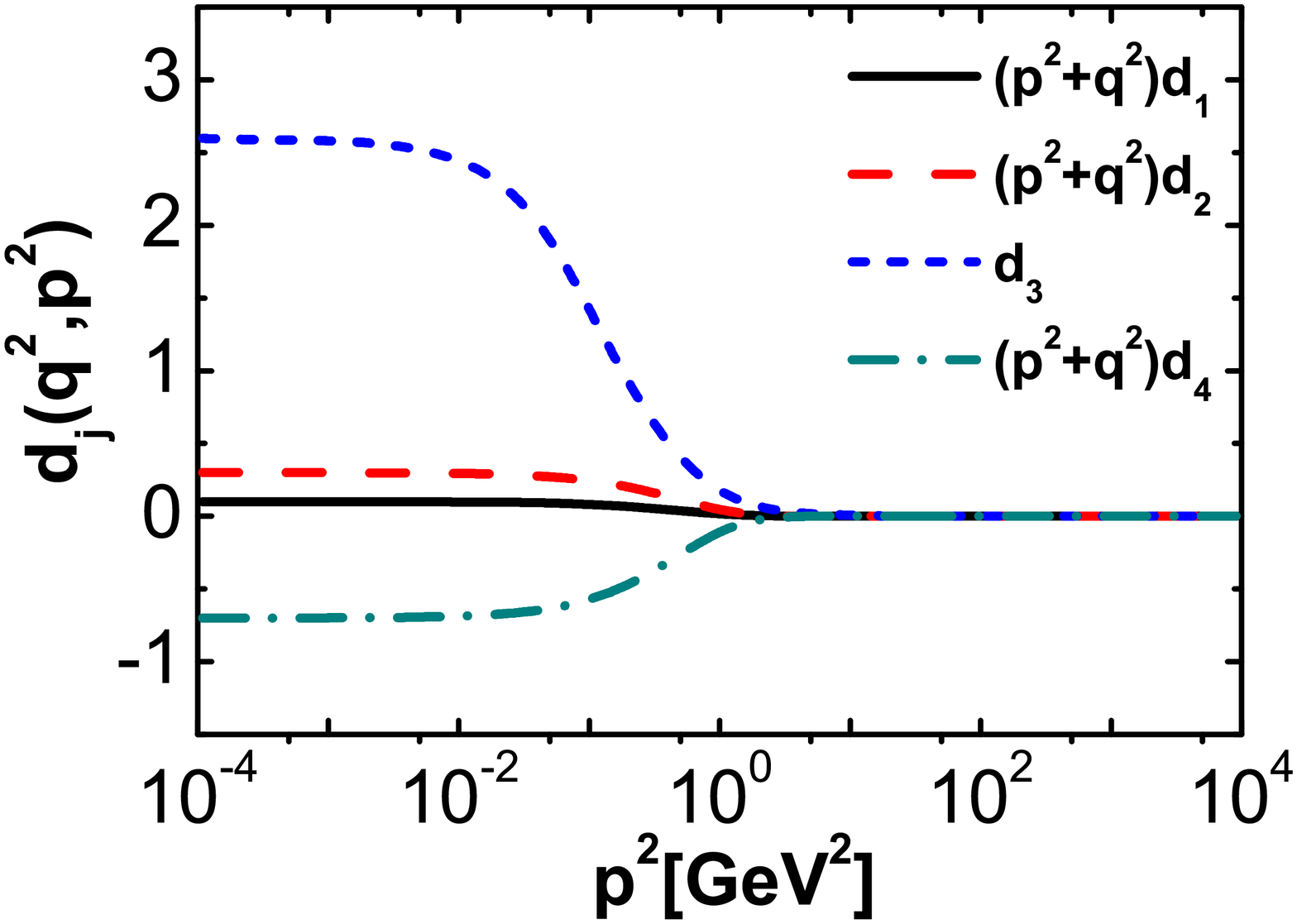}
\vspace*{-6mm}
\includegraphics[width=0.42\textwidth]{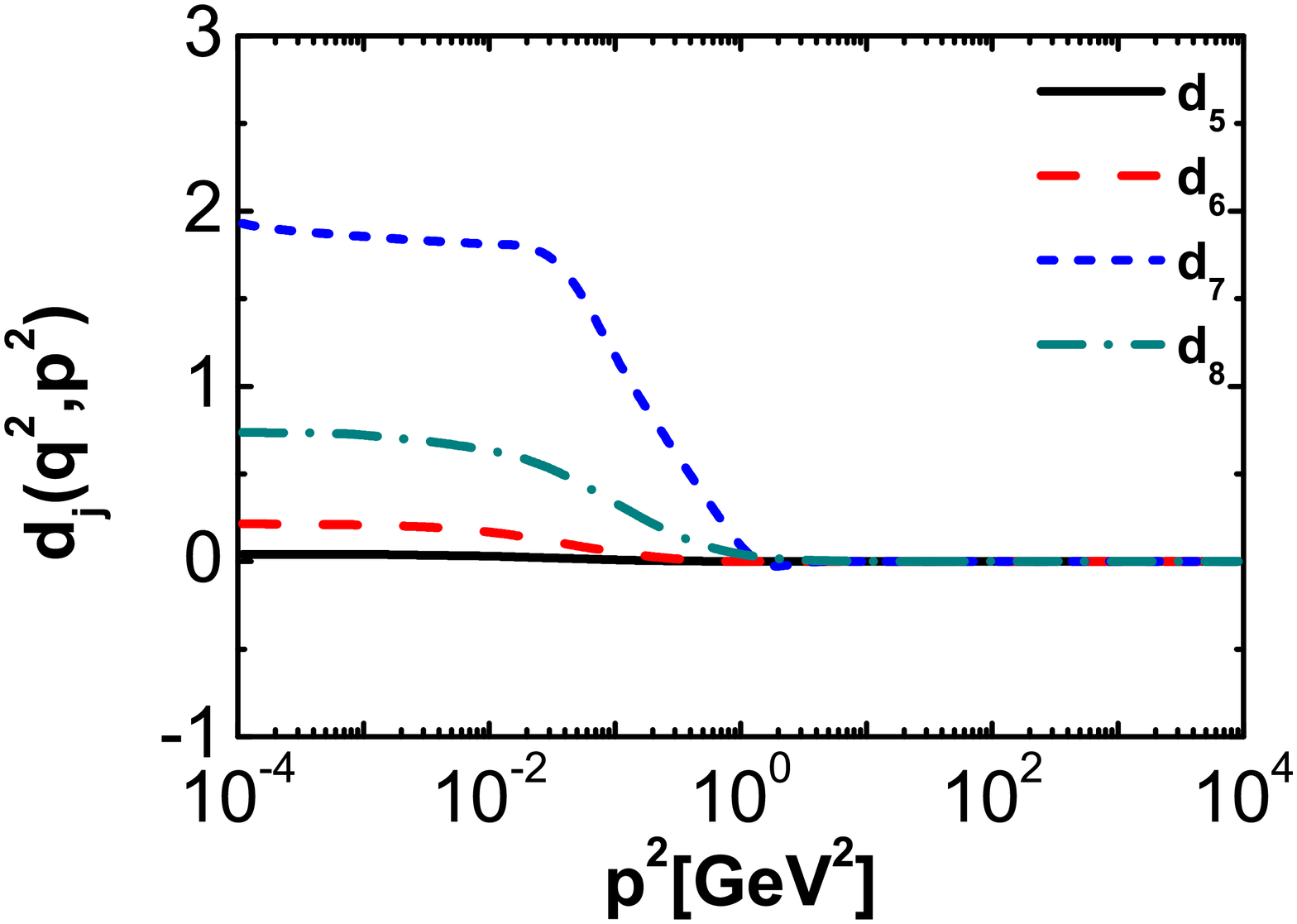}
\end{center}
\vspace*{-2mm}
\caption{Calculated momentum dependence of the transverse structures's coefficients of the quark-photon vertex.}
\label{fig:qqp2}
\end{figure}

The obtained results of the coefficients $c_{i}^{}(q^2 , p^2) \;\; (i = 1,2,3,4)$ and $d_{i}^{}(q^2, p^2) \; \; (i = 1,2, \cdots, 8)$
at the symmetric momentums ($p^2=q^2$) are shown in Fig.~\ref{fig:qqp1} and Fig.~\ref{fig:qqp2}.

The upper panel of Fig.~\ref{fig:qqp1} shows the obtained dressing functions of the longitudinal part without  considering the constraint of the WTI and the comparison with those in RL approximation~\cite{AW:2018CPC}.
It is clear that the $c_{4}^{}$ deviates from zero very slightly, which means that the present truncation scheme does not break the symmetry greatly, even if we take the approximation in Eq.~(\ref{app}).
Comparing the RL approximation results with the ``full" ones, one can know that the RL approximation enlarges
the contributions of the $L^{1}$ and $L^{3}$ terms, but suppresses that of the $L^{2}$ term.
So does the lower panel in Fig.~\ref{fig:qqp1} but for the ones with the constraint of the WTI,
from which one can notice apparently that the presently obtained results of the momentum dependence of the $\Sigma_{A}^{}$ and  $\Delta_{B}^{}$ are quite close to those from the RL approximation ($c_{1}^{\prime}$ and $c_{3}^{\prime}$)~\cite{AW:2018CPC}.
While the  behaviors of  $\Delta_{A}^{}$ and $c_{2}^{\prime}$ are different.
The $c_{2}^{\prime}$ is close to 0 at low momentum,
however the $\Delta_{A}$ holds negative value with quite large amplitude in the IF region and increases monotonously to $0$ as the momentum increases.
Such a behavior of $\Delta_{A}^{}$ is consistent with that given in
lattice QCD simulations~\cite{LS:2018} qualitatively.
This shows that, before the completely symmetry-preserved truncation is build, the WTI constraints taken here are necessary for quark photon vertex.
Comparing the results shown in the two panels carefully, one can recognize that the constraint of the WTI enhances the contributions of all the $L^{1}$, $L^{2}$ and $L^{3}$ terms.

Numerical data indicate that the momentum dependence of the coefficients of the Lorentz structure
$T^{1}$, $T^{2}$ and $T^{4}$ are similar to that of the corresponding term of the quark-gluon interaction vertex,
 i.e., they involve $1/(p^{2} + q^{2})$ divergence as $p^{2} + q^{2}$ approaches to $0$.
Then in Fig.~\ref{fig:qqp2} we show the  behaviors of the coefficients $d_{1}^{}$, $d_{2}^{}$
and $d_{4}^{}$ with a factor $(p^{2} + q^{2})$ being multiplied.
After comparing the  behaviors of the coefficients $d_{i}^{} \;\; (i =1,2, \cdots, 8)$
shown in Fig.~\ref{fig:qqp2}, one can notice that the contribution of the Lorentz structure $T^{3}$,
$T^{7}$ and $T^{8}$ are dominant, which is similar to the quark-gluon interaction vertex.
These quite large contributions of the transverse structure of the quark-photon vertex will lead to
observable signals.
Besides, the transverse part of the quark-photon vertex with different tensor structures has been studied in the RL approximation (e.g., Ref.~\cite{AW:2018CPC}). After comparison, we find that the $\tau_{3}$ and $\tau_{4}$ in Ref.~\cite{AW:2018CPC} are respectively the $d_{5}$, $d_{8}$ in this work, and the corresponding coefficients are qualitatively coincident with each other. Especially, the magnitude of  $d_5$ is always smaller compared to the other components.

\vspace*{-5mm}

\subsection{Anomalous magnetic moment}

\begin{figure}[hbt]
\centerline{\includegraphics[width=0.46\textwidth]{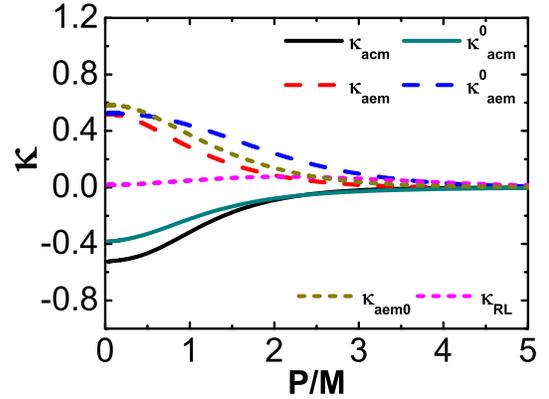}}
\vspace*{-3mm}
\caption{Calculated momentum dependence of the ACM and AEM of a dressed-quark (The unit $M$ is the mass of the dressed-quark at zero momentum). The $\kappa_{acm}^{}$, $\kappa_{aem}^{}$ ($\kappa_{aem0}^{}$) is the ACM, AEM obtained with the presently proposed truncation scheme and with (without) the constraint of the WTI,  
respectively.
The $\kappa^{0}_{acm}$ and $\kappa^0_{aem}$ are those obtained from the CLRQ vertex~\cite{CLR}
and the interaction model given in Ref.~\cite{MR1997}.
The $\kappa_{RL}^{}$ is the AEM obtained from the Rainbow-Ladder approximation.}
\label{fig:ACM}
\end{figure}

We then calculate the quark's anomalous chromo-magnetic moment (ACM) and anomalous electro-magnetic moment (AEM) in the same way as that in  Ref.\cite{CLR}, namely
\begin{equation} \label{acm1}
\kappa(\varsigma) = \frac{2\varsigma F_{2}^{} + 2 \varsigma^{2} F_{4} + \Lambda_{\kappa}^{}(\varsigma)} {a_{1}^{} + F_{1}^{} - \Lambda_{\kappa}^{}(\varsigma)} \, ,
\end{equation}
where $\Lambda_{\kappa}^{}(\varsigma) = 2\varsigma^{2} a_{2} - 2\varsigma a_{3}^{} - \varsigma F_{5}^{} -\varsigma^{2} F_{7}$ with the $a_{i}^{}$ and $F_{i}^{}$ being evaluated at $q^{2} = p^{2} = M(p^{2})
=\varsigma^{2}$.
In which the functions $a_{1}^{}$, $a_{2}^{}$ and $a_{3}^{}$ are just correspondingly the coefficients of
the $L^{1}$, $L^{2}$ and $L^{3}$ defined in Eq.~(\ref{tensors1L-photon}), namely the longitudinal structure of the vector vertex.
The $F_{i}^{}$($i=1,2, \cdots, 8$) are the coefficients of the tensors redefined as
\begin{eqnarray}\label{acm2}
\begin{split}
\Gamma^{\gamma,T}_\mu(q,p)= & \; \gamma^{T}_{\mu} F_{1}^{} + \sigma_{\mu\nu}^{} k_{\nu}^{} F_{2}^{}
+ T_{\mu\rho}^{} \sigma_{\rho\nu}^{} l_{\nu}^{} l \cdot k F_{3}^{}  \\
&\; + [l^{T}_{\mu}\gamma \cdot k + i \gamma^{T}_{\mu} \sigma_{\nu\rho}^{} l_{\nu}^{} k_{\rho}^{}]F_{4}^{}
-i l^{T}_{\mu} F_{5}^{} \qquad \\
&\; + l^{T}_{\mu} \gamma\cdot k l\cdot k F_{6}^{} - l^{T}_{\mu} \gamma\cdot l F_{7}^{}  \\
&\; + l^{T}_{\mu} \sigma_{\nu\rho}^{} l_{\nu}^{} k_{\rho}^{} F_{8}^{} \, ,
\end{split}
\end{eqnarray}
where $\sigma_{\mu \nu}^{} = \frac{i}{2}(\gamma_\mu\gamma_\nu-\gamma_\nu\gamma_\mu)^{}$; $l$ and $k$ are respectively the $l=(p+q)/2$ and $k=q-p$.
The conversion from the coefficients $a_{i}^{}\;\;(i=1,2,3,4)$ and $b_{j}^{}\;\; (j=1,2,\cdots,8)$ of the standard tensors given in Section~\ref{Subsec:Solutions-qg-Vetex} to the coefficients $F_{i}^{}$ here is the same as that described in Section~\ref{Sec:Algorithm}.
The obtained quark's anomalous chromo- and electro-magnetic moment
from the the presently proposed truncation scheme
and the comparison with that given by the CLRQ vertex is illustrated in Fig.\ref{fig:ACM}.

Fig.~\ref{fig:ACM} manifests apparently that 
the sign of the ACM obtained in our present work is the same as that given with the CLRQ vertex.
Such a negative ACM agrees with the general requirement mentioned in Ref.~\cite{BCPQR},
which comes from the monotonically decreasing dressed-quark mass function, and matches the constraints of perturbative QCD. While the positive AEM also matches the constraints of perturbative QCD.

Looking through the figure more cautiously, one can observe that the presently obtained vertex
leads to a larger amplitude for the ACM than the CLRQ vertex,
while the RL approximation gives only nearly zero anomalous magnetic moment.
These results reveal that our presently obtained quark-gluon interaction vertex
carries more DCSB effect than the CLRQ model and the RL approximation,
even though the DCSB effect demonstrated by the quark propagator, i.e., quark condensate, is similar.
 Moreover, if one does not take the constraint of the WTI on the quark-photon vertex into account,
the obtained result of the anomalous electric-magnetic moment is close that with WTI constraint, but slightly enhanced in the infrared region.

\section{Summary}       \label{Sec:Summary}

We construct a scheme that couples the DSEs of the quark propagator and the quark-gluon interaction vertex in this paper.
In this scheme, the physical coupling strength and gluon propagator from ab initio computation can be implemented directly, which then builds a connection  from the QCD Lagrangian to the phenomenology numerically.
We study the quark-gluon interaction vertex which includes all the 12 Lorentz structures,
and give the  behaviors of the vertex with respect to the momentums (including the angle between them) and the current quark mass.
Our direct computation gives consistent result with those from phenomenological ansatz and lattice QCD simulations.
We also give a parameterized expression for the obtained quark-gluon vertex
to the convenience of using it elsewhere.
We  show furthermore the practicality of this scheme by combining it with the BSEs
and producing reasonable quark-photon interaction vertex and anomalous chromo- and electro-magnetic moments of quark.
This may shed light on building the realistic scheme beyond RL approximation to study hadron properties.
The work is under progress.

\medskip

The work was supported by the National Natural Science Foundation of China under Contracts No.\ 11435001,
No. 11775041 , and the National Key Basic Research Program of China under Contract No.\ 2015CB856900.
FG was also supported by the Spanish MEyC, under grants FPA2017-84543-P and SEV-2014-0398; Generalitat Valenciana under grant Prometeo II/2014/066.
We are grateful for the discussions and comments from A.C. Aguilar,   A. Bashir, L. Chang, J. Papavassiliou,  C.D. Roberts,  and J. Rodriguez-Quintero.
%


\section*{Appendix}
\medskip
We list here the standard tensor structures of the full quark-gluon interaction vertex.
Those of the longitudinal part are given in Eqs.(\ref{tensors11}) and the transverse ones are listed in Eqs.(\ref{tensors12}):
\begin{eqnarray}\label{tensors11}
\begin{split}
&L^{1}_{\mu}(t,k) =\gamma_{\mu}^{} \, ,  \qquad \qquad L^{2}_{\mu}(t,k)=\frac{1}{2}t_{\mu} \slashed{t} \, , \\[1mm]
&L^{3}_{\mu}(t,k) = -i t_{\mu} I_{D}^{} \, , \qquad L^{4}_{\mu}(t,k) = -\frac{i}{2}(\gamma_{\mu}^{} \slashed{t}
- \slashed{t} \gamma_{\mu}^{} ) \, , \qquad
\end{split}
\end{eqnarray}
the transverse part is,
\begin{eqnarray}\label{tensors12}
\begin{split}
&T^{1}_{\mu}(t,k) = \frac{i}{2}(k^{2} t_{\mu} - t\cdot k k_{\mu}^{} ) \, , \\
&T^{2}_{\mu}(t,k) = \frac{1}{2}(k^{2} t_{\mu} - t\cdot k k_{\mu}^{} ) \slashed{t} \, , \\
&T^{3}_{\mu}(t,k) = k^{2} \gamma_{\mu} - k_{\mu} \slashed{k} \, , \\
&T^{4}_{\mu}(t,k) = \frac{i}{8}(k^{2} t_{\mu} - t\cdot k k_{\mu})( \slashed{t} \slashed{k} - \slashed{k} \slashed{t}) \, , \\
&T^{5}_{\mu}(t,k) = \frac{i}{2}(\slashed{k} \gamma_{\mu}^{} - \gamma_{\mu}^{} \slashed{k} ) \, , \\
&T^{6}_{\mu}(t,k) = t\cdot k\gamma_{\mu}^{} - \slashed{k} t_{\mu} \, ,  \\
&T^{7}_{\mu}(t,k) = \frac{i}{2} t\cdot k (t_{\mu}^{} - \gamma_{\mu}^{} \slashed{t})
 +\frac{i}{4} t_{\mu}^{} (\slashed{k} \slashed{t} - \slashed{t} \slashed{k}) \, ,  \\
&T^{8}_{\mu}(t,k) = \frac{1}{2}( \slashed{k} t_{\mu}^{} - \slashed{t} k_{\mu}^{})
+\frac{\gamma_{\mu}^{}}{4} (\slashed{k} \slashed{t} - \slashed{t} \slashed{k} ) \, ,
\end{split}
\end{eqnarray}
where $I_{D}^{}$ is the $4\times 4$ identity matrix in Dirac space, $k_\mu=p_\mu-q_\mu$ and $t_\mu=p_\mu+q_\mu$.

\medskip

The tensor set whose constituents are orthogonal with each other are listed in Eqs.~(\ref{tensors21}) and (\ref{tensors22}), where the longitudinal part reads
\begin{eqnarray} \label{tensors21}
\begin{split}
& L'^{1}_{\mu} (t,k) = \frac{ \slashed{k} k_{\mu}^{}}{k^{2}} \, ,  \\
& L'^{2}_{\mu} (t,k) = \frac{t\cdot k \slashed{t} - \frac{(t\cdot k)^{2} \slashed{k}}{k^{2}}}{2k^{2}} k_{\mu}^{} \, , \\
& L'^{3}_{\mu} (t,k) = -i \frac{t\cdot k k_{\mu}^{}}{k^{2}} \, ,  \\
& L'^{4}_{\mu} (t,k) = i \frac{( \slashed{t} \slashed{k} - \slashed{k} \slashed{t}) k_{\mu}^{}}{2k^{2}} \, ,
\end{split}
\end{eqnarray}

The transverse part is
\begin{eqnarray}\label{tensors22}
\begin{split}
T'^{1}_{\mu}(t,k) = & \; \gamma_{\mu}^{} - \frac{ \slashed{k}   k_{\mu}^{}}{k^{2}} \, , \\
T'^{2}_{\mu}(t,k) = & \; 3( t_{\mu}^{} - \frac{t\cdot k k_{\mu}^{}}{k^{2}})( \slashed{t} - \frac{ \slashed{k} t\cdot k}{k^{2}}) \\
 & - (\gamma_{\mu}^{} - \frac{ \slashed{k} t\cdot k}{k^{2}})(t^{2} - \frac{(t\cdot k)^{2}}{k^{2}}) \, , \\
T'^{3}_{\mu}(t,k) = &\; t\cdot k( t_{\mu} - \frac{t\cdot k k_{\mu}^{}}{k^{2}}) \slashed{k} \, , \\
T'^{4}_{\mu}(t,k) = &\; \frac{1}{2}( \slashed{k} t_{\mu}^{} - \slashed{t} k_{\mu}^{})
                           +\frac{\gamma_{\mu}^{}}{4}(\slashed{k} \slashed{t} - \slashed{t} \slashed{k}) \, , \\
T'^{5}_{\mu}(t,k) = & -i(t_{\mu}^{} - \frac{t\cdot k_{\mu}^{}}{k^{2}}) \, ,  \\
T'^{6}_{\mu}(t,k) = & -it \cdot k [(\gamma_{\mu}^{} - \frac{ \slashed{k} k_{\mu}^{}}{k^{2}})( \slashed{t} - \frac{\slashed{k} t\cdot k}{k^{2}}) \qquad \\
& - (\slashed{t} - \frac{ \slashed{k} t\cdot k}{k^{2}})(\gamma_{\mu}^{} - \frac{\slashed{k} k_{\mu}^{}}{k^{2}})] \, ,  \\
T'^{7}_{\mu} (t,k) = & -i(t^{2} - \frac{(t\cdot k)^{2}}{k^{2}})(\gamma_{\mu}^{} \slashed{k} - \slashed{k} \gamma_{\mu}^{}) \\
& + 2i(t_{\mu}^{} - \frac{t\cdot k k_{\mu}^{}}{k^{2}})(\slashed{t} \slashed{k} - t\cdot k) \, ,  \\
%
%
T'^{8}_{\mu} (t,k) = & -i(t_{\mu}^{} - \frac{t\cdot k k_{\mu}^{}}{k^{2}})(\slashed{t} \slashed{k} - t\cdot k) \, .
\end{split}
\end{eqnarray}
Such an expression of the transverse part of the orthogonalized tensors
are from the orthogonalized vector basis of the vector meson~\cite{QCLR2,VPDA}.
After changing the form of the $L^{i}_{\mu}$, the 12 independent orthogonalized tensors can be built.



\end{document}